\documentclass[runningheads]{svmult}

\usepackage{graphicx}  

\newcommand{\tit}[1]{} 
\newcommand{\disstit}[1]{`#1',}
\newcommand{\pag}[2]{pp.~#1}
\newcommand{\art}[5]{{\em #1} {\bf #2}, #3 #5} 
\newcommand{\bib}[1]{\bibitem{#1}}
\begin{document}
\unitlength=1.0cm
\title*{Critical Discussion of ``Synchronized Flow'',\protect\newline 
Simulation of Pedestrian Evacuation, and\protect\newline 
Optimization of Production Processes}
\toctitle{Critical Discussion of ``Synchronized Flow'', Simulation of\protect\newline 
Pedestrian Evacuation, and Optimization of Production Processes}
%
%
\titlerunning{``Synchronized Flow'', Pedestrian Evacuation, and Optimal Production}
%
\author{Dirk Helbing\inst{1}
\and Ill\'{e}s J. Farkas\inst{2}
\and Dominique Fasold\inst{1}
\and Martin Treiber\inst{1}
\and Tam\'{a}s Vicsek\inst{2}}
\authorrunning{Dirk Helbing et al.}
%
%
\institute{Institute for Economics and Traffic, Dresden University of Technology,
Andreas-Schubert-Str. 23, D-01062 Dresden, Germany
\and 
Department of Biological Physics, E\"otv\"os University, Budapest,
P\'azm\'any P\'eter S\'et\'any 1A, H-1117 Hungary}

\maketitle              

\begin{abstract}
We critically discuss the concept of ``synchronized flow'' from a
historical, empirical, and theoretical perspective. Problems related to the
measurement of vehicle data are highlighted, and questionable interpretations
are identified. Moreover, we propose a quantitative and consistent theory 
of the empirical findings based on a phase diagram of congested traffic states, which is universal
for all conventional traffic models having the same instability diagram and a 
fundamental diagram. New empirical and simulation data supporting this approach
are presented as well. We also give a short overview of the various phenomena
observed in panicking pedestrian crowds relevant from the point of 
evacuation of buildings, ships, and stadia. 
Some of these can be applied to the optimization of production processes,
e.g. the ``slower-is-faster effect''.
\end{abstract}

\section{Freeway Traffic: ``Synchronized flow'', ``Pinch Effect'', and Measurement Problems}

\subsection{What is new about ``synchronized flow?''}

\begin{figure}[htbp]
\begin{center}
\includegraphics[width=6cm]{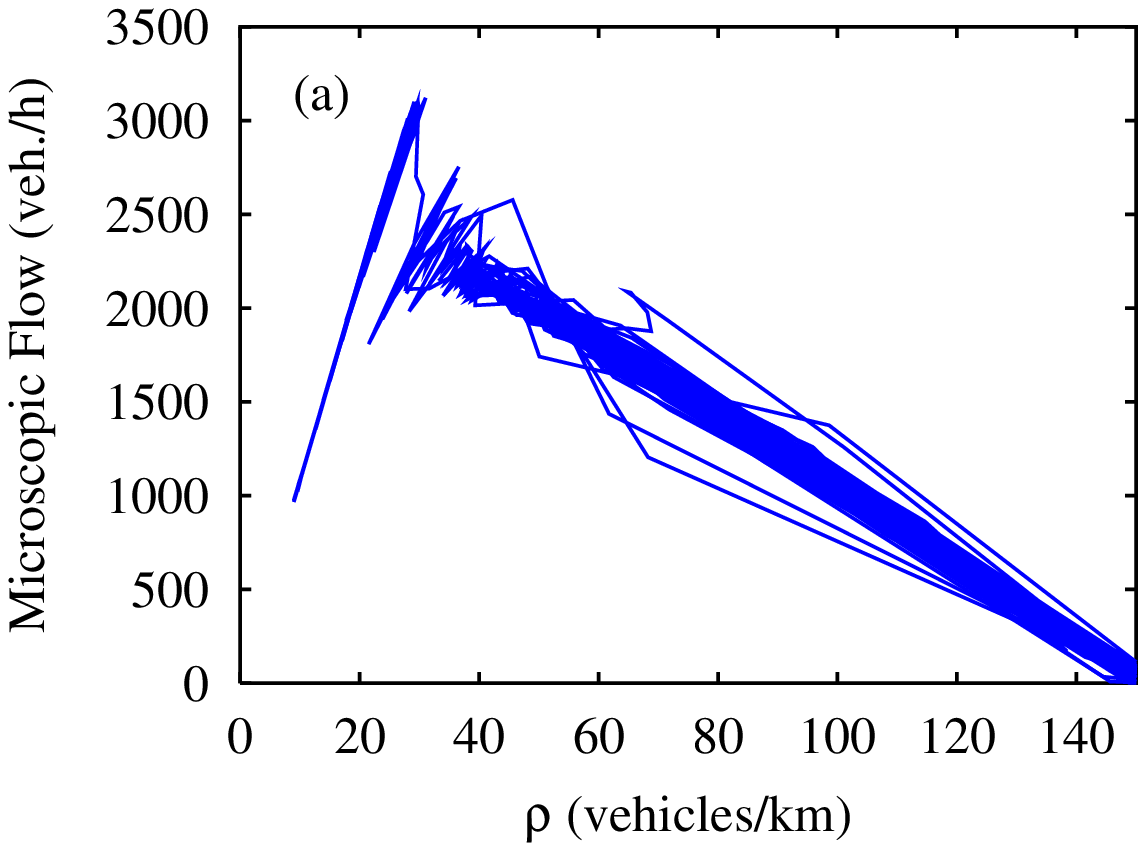}
\includegraphics[width=6cm]{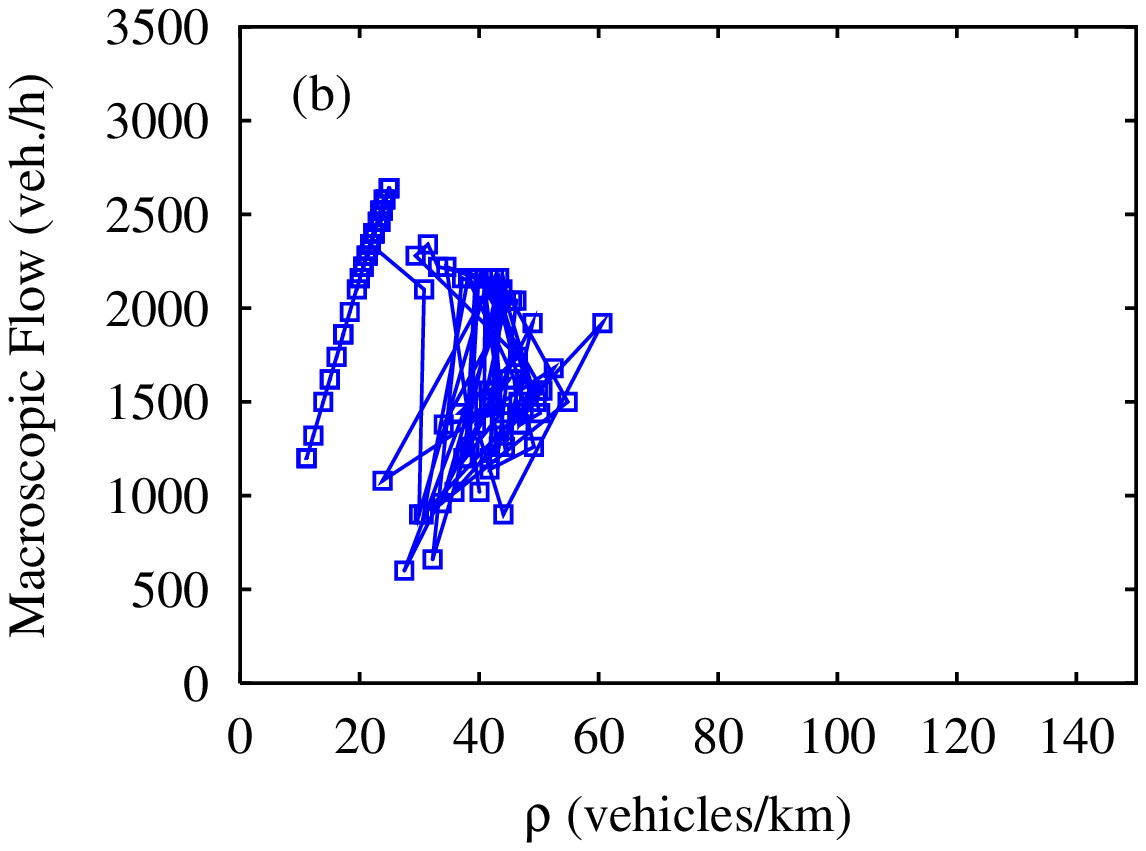}
\end{center}
\caption[]{(a) Flow-density relation for narrow moving jams simulated with a microscopic
traffic model. (b) The aggregated (1-minute) data corresponding to the narrow moving
jams displayed in (a) show a wide scattering and erratic movement 
in the flow-density plane. By mistake, this is used to characterize 
``synchronized flow''.\label{Fig1}}
\end{figure}

\begin{figure}[htbp]
\begin{center}
\begin{picture}(12,9.7)(-1.6,0)
\put(0.15,10){\includegraphics[width=5.3\unitlength,height=8.4\unitlength,
angle=-90]{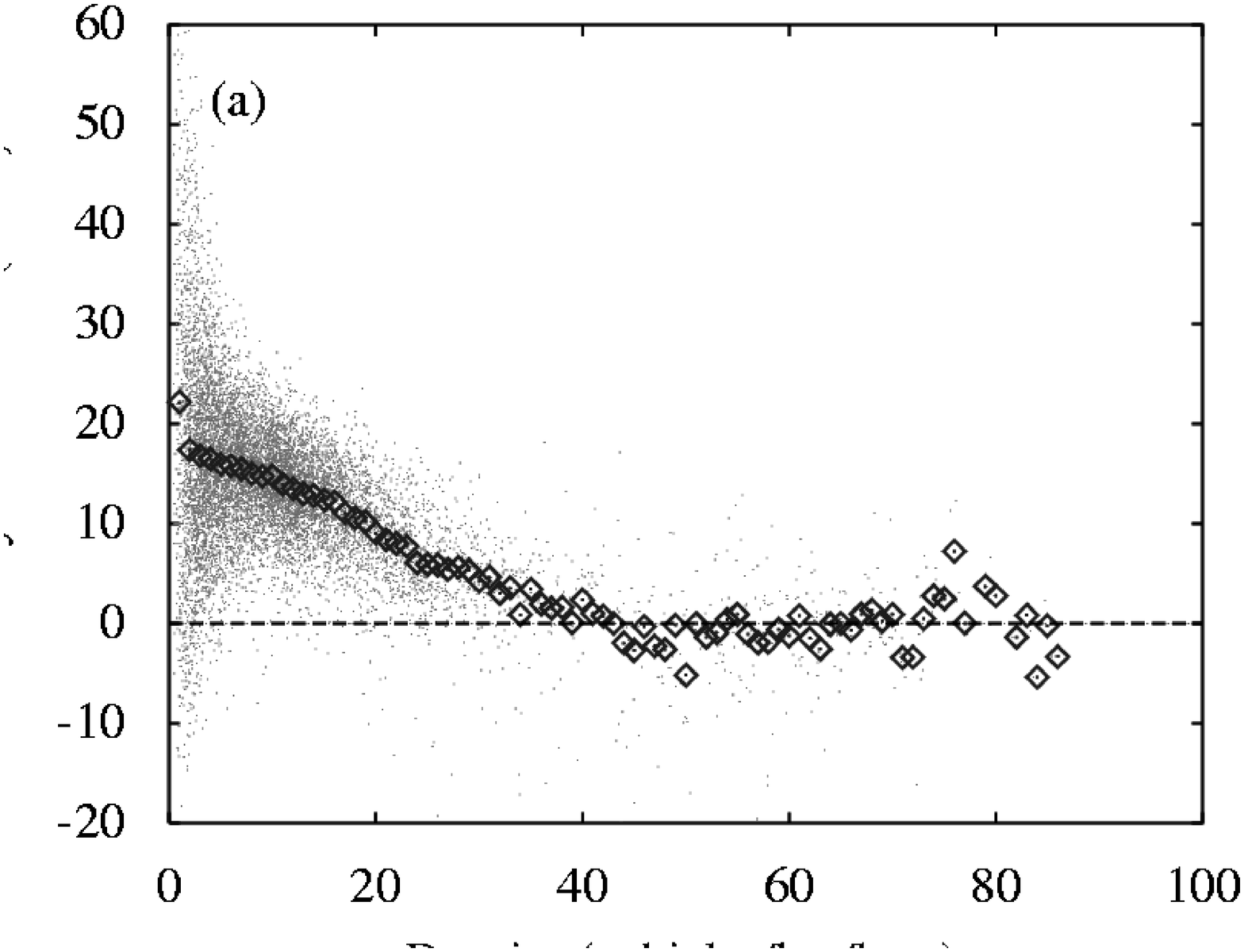}} 
\put(-0.4,4.5){\includegraphics[height=5\unitlength,angle=-90]{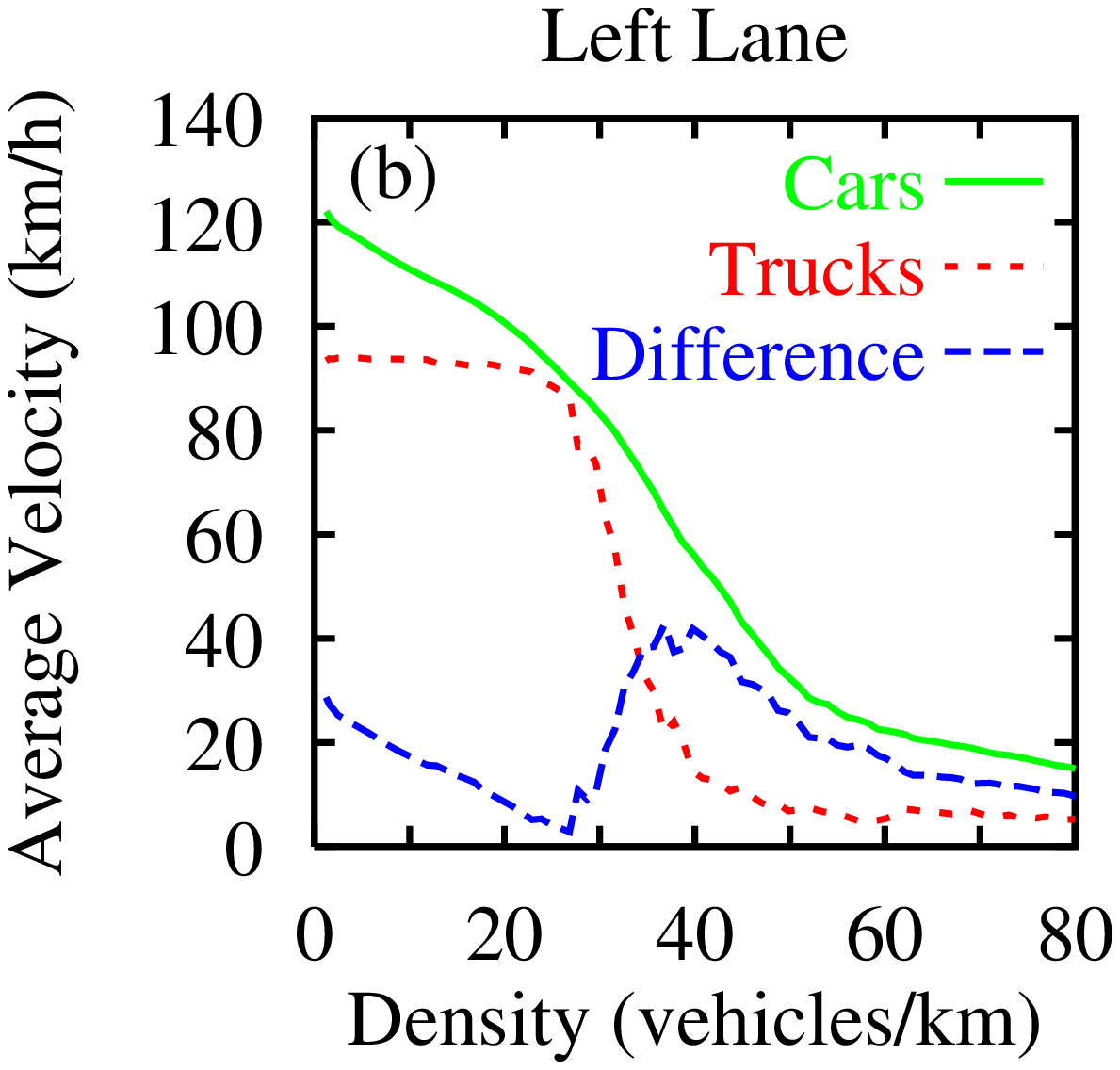}}
\put(3.6,4.5){\includegraphics[height=5\unitlength,angle=-90]{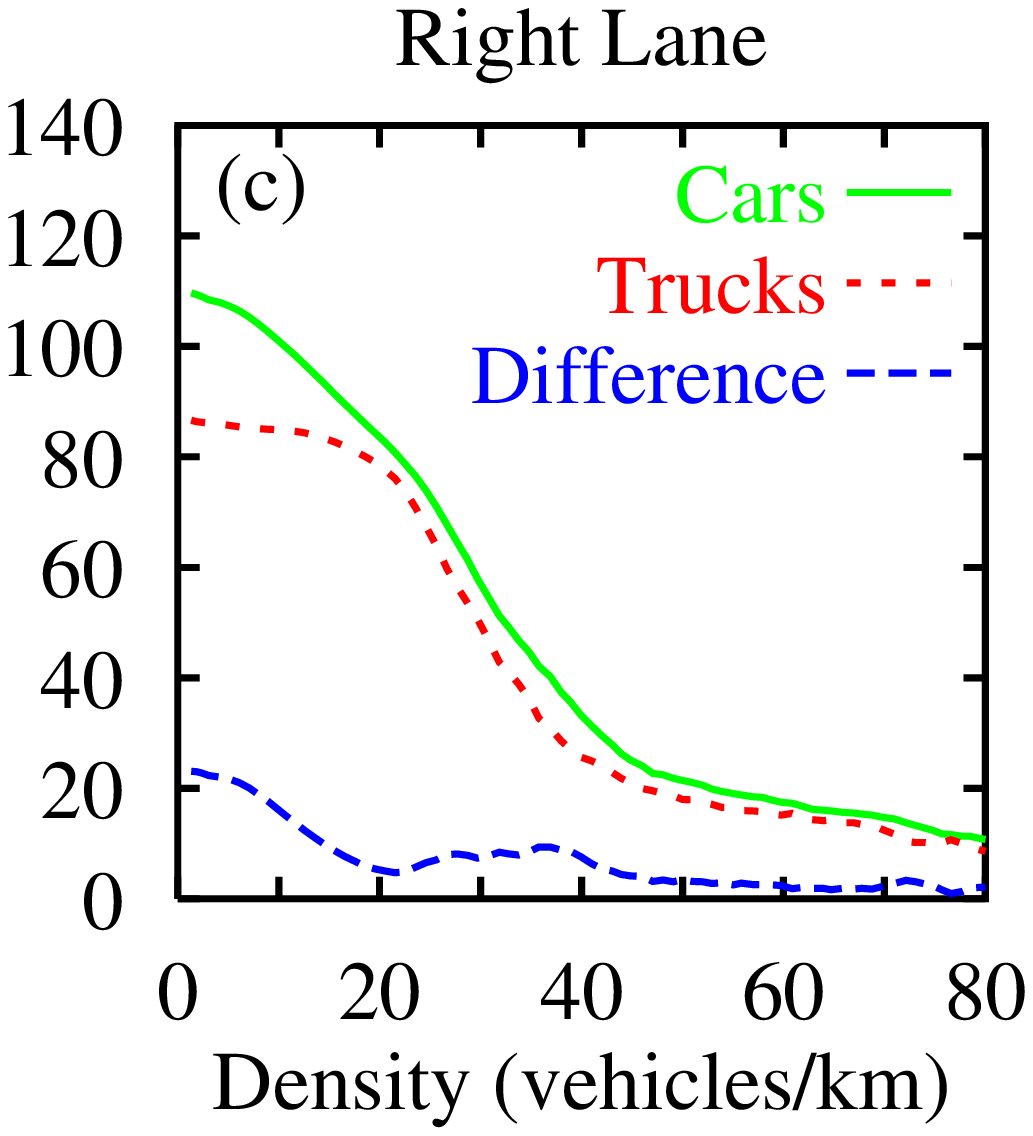}}
\end{picture} 
\end{center}
\caption[]{(a) The difference of the average velocity in the left and the right lane
vanishes at densities above 30 vehicles per kilometer, corresponding to
a macroscopic synchronization of the speeds \cite{Hel97a}. 
(b), (c) The difference in the empirically determined velocities of cars and trucks,
however, show that overtaking maneuvers continue to exist even at higher 
densities \cite{HelHu98}.
\label{Fig2}}
\end{figure}

Congested traffic has been investigated for many decades because of its complex phenomenology. 
Therefore, Kerner and Rehborn have removed the data belonging to wide 
moving jams (see MLC and TSG in Figs.~\ref{Fig9} and \ref{Fig10}) and found
found that the remaining data of congested traffic
data still displayed a wide and two-dimensional scattering \cite{KerRe96b}, 
see Fig.~\ref{Fig4}(c). By mistake (see Figs.~\ref{Fig1}, \ref{Fig4}(c), and Sec.~\ref{widescat}),
they concluded that all models assuming a fundamental diagram were wrong
and defined a new state called {\em ``synchronized flow''} (``synchronized'' because of the typical
synchronization among lanes in congested traffic, see Fig.~\ref{Fig2}(a),
and ``flow'' because of flowing in contrast to standing traffic in fully developed jams).
Since then, Kerner suggests to classify {\em three phases}:
(1) {\em free flow,}
(2) {\em ``synchronized flow''}, and 
(3) {\em wide moving jams} (i.e. {\em moving localized clusters} whose width in longitudinal
direction is considerably higher than the width of the jam fronts). In some
applied empirical studies, however, Kerner {\em et al.} additionally 
distinguish a fourth state of stop-and-go traffic \cite{FOTO,KerReAl00}.
\par
{\em Free flow} is characterized by the average desired velocity $V_0$ and,
therefore, by a strong correlation and quasi-linear relation $Q \approx \rho V_0$
between the local flow $Q$ and the {\em local} density $\rho$ \cite{NeubSaScSc99}.
It is also well-known that {\em wide moving jams} propagate with constant form and
(phase) velocity $C\approx -13$km/h  \cite{MikaKrYu69,KerRe96a,CasMa99}. 
Kerner found that this propagation is not 
affected by bottlenecks or the presence of ``synchronized flow''.
Moreover, he showed that the outflow $Q_{\rm out}$ from wide jams is a 
self-organized traffic constant as well \cite{KerRe96a,Kerner1}.
In contrast to wide moving jams, the flow inside of {\em ``synchronized flow''} remains finite,
and its downstream front is normally 
fixed at the location of some bottleneck, e.g. an on-ramp. Therefore, ``synchronized flow'' 
basically agrees with 
previous observations of queued or congested traffic (see, e.g., Refs.
\cite{Per86,Bank90,Bank91a} and the references therein). 
In his patent \cite{KerKirReh}, Kerner
applies the queuing theory himself, which goes back to
the fluid-dynamic traffic model by Lighthill and Whitham \cite{LighWh55}. 
\par
The synchronization of the
average velocities among neighboring lanes has been already described by Koshi {\em et al.}
\cite{KosIwOh83} (but see also Refs. \cite{MikaKrYu69,EdiFo58,ForMuSi67}).
It is true on a {\em macroscopic} level for {\em all} forms of congested traffic 
including wide moving jams. Simulations have shown that this is
a result of lane changes \cite{Lee3LeKi98}, 
while the assumed reduction in the lane changing rate \cite{KerRe97} occurs only after 
the speeds in neighboring lanes have been successfully balanced \cite{ShvHe99}.
On a {\em microscopic} scale,
over-taking maneuvers continue to exist almost at all densities \cite{HelHu98},  see Figs.~\ref{Fig2}(b), (c). 
Nevertheless, the probability of lane changes drops considerably with increasing density,
when most of the road is used up by the vehicles' safety headways \cite{HelHu98},
but not in the postulated $Z$-shaped way \cite{Kerner3}. Due to the reduced opportunities
for overtaking and the related bunching of vehicles, the velocity variance 
goes down with increasing density as well \cite{KerRe97,Hel97a,Hel97c}. 
\par
The transition between free and congested traffic is of hysteretic nature,
i.e. the inverse transition occurs at a lower density and a higher average 
velocity. This has been known for a long time \cite{TreitMy74,Pay84,Hal187}.
Kerner specifies that the transition is usually caused by a localized
and short perturbation of traffic flow that starts downstream of the 
on-ramp and propagates upstream with a velocity of about $-13$\,km/h.
As soon as the perturbation passes the on-ramp, it triggers the breakdown 
which spreads upstream in the course of time. The congested state can then persist for
several hours \cite{KerRe97}. 
\par
\begin{figure}[htbp]
\vspace*{-3mm}
\begin{center}\hspace*{-2mm}
\includegraphics[width=7cm]{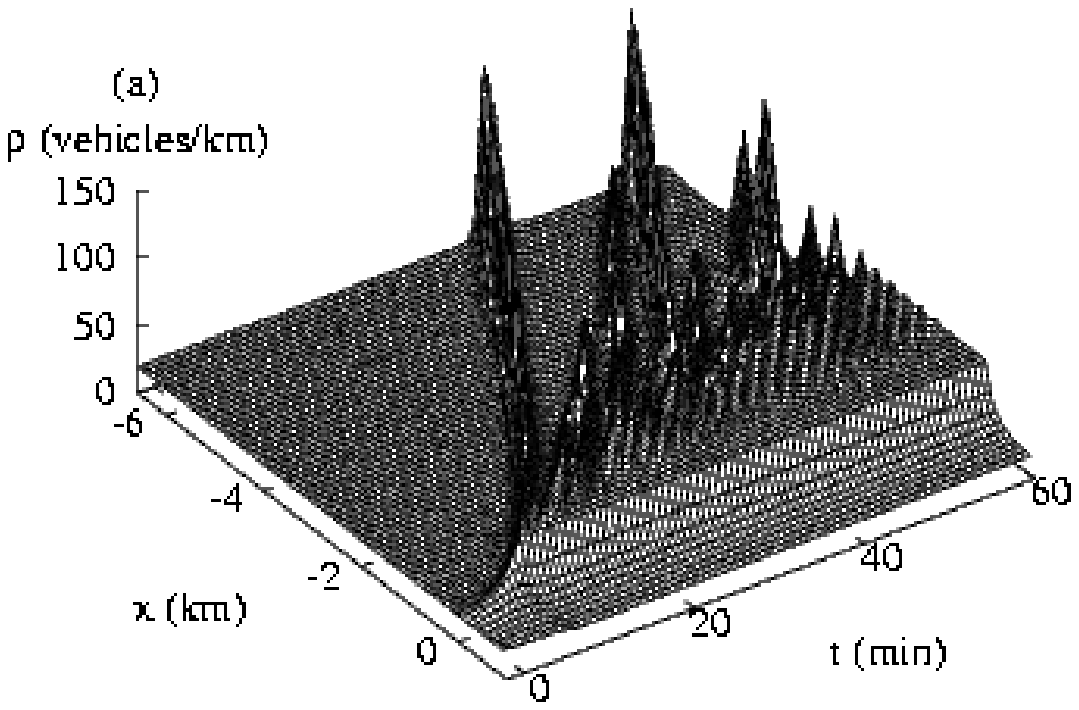}\hspace*{-6mm}
\includegraphics[width=6cm]{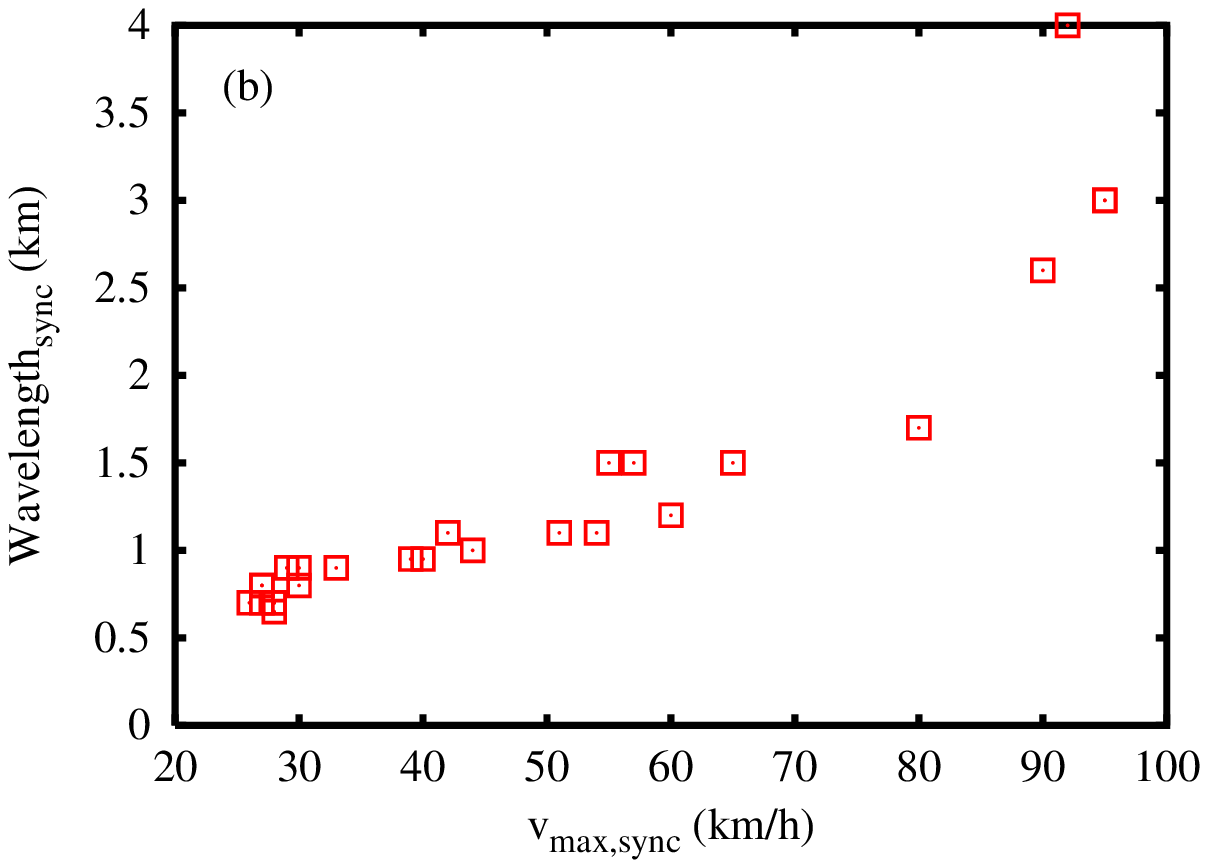}
\end{center}
\caption[]{(a) Simulation of the pinch effect with a deterministic microscopic model 
showing stable traffic at low and
high densities, linearly unstable traffic at medium densities, and metastable traffic
in between. The spatio-temporal density plot illustrates
the breakdown to homogeneous congested traffic (HCT) upstream of a bottleneck,
emerging oscillating congested traffic (OCT) further upstream, and a few stop-and-go waves
(TSG) emanating from this region. The conditions for this 
spatial coexistence of congested traffic states are as follows \cite{Review,TreHe99b}: 
The density in the congested region immediately 
upstream of the bottleneck should be in the linearly unstable, but convectively stable
range, where perturbations are convected away in 
upstream direction \cite{Mann90,CrosHo93}.
In this case, traffic flow will appear stationary and homogeneous close to the bottleneck,
but small perturbations will grow as they propagate upstream in the congested
region starting at the bottleneck. If the perturbations propagate faster than
the congested region expands, they will reach the area of free traffic upstream of the
bottleneck. During rush hours, it is quite likely that this free flow is in the metastable
range between the free and linearly unstable density region.
Consequently, sufficiently large perturbations
can trigger the formation of jams, which continue travelling upstream, while
small perturbations are absorbed. (b) The wavelength (average distance between
density maxima) determined from (a) qualitatively displays the empirical increase with the
vehicle velocity observed by Kerner \cite{Ker98a}.\label{Fig3}}
\end{figure}

Moreover, Kerner and Rehborn distinguish three types of ``synchronized flow'' \cite{KerRe96b},
which may be short-lived:
(i) Stationary and homogeneous states where both the average
speed and the flow rate are stationary 
(see, e.g., also Refs. \cite{Hal1Ag91,PerYaBr98,Wes98}. Later on,
we will these {\em ``homogeneous congested traffic''} (HCT) \cite{HelHeTr99}.\\
(ii) States where only the average vehicle speed is stationary, 
named {\em ``homogeneous-in-speed states''} (see also 
Refs. \cite{Ker98b,Lee3LeKi00}).
We interpret this state as {\em ``recovering traffic''} \cite{Review}, 
as it bears several signatures of free traffic and mostly appears downstream 
of bottlenecks with congested traffic.\\
(iii) Non-stationary and non-homogeneous states
(see also Refs. \cite{Ker98b,CasBe99,TreHeHe00}).
For these, we will use the term {\em ``oscillating congested traffic''} (OCT)
\cite{HelHeTr99}.\\
At least types (i) and (iii) are characterized by a considerably scattering and erratic
change of the flow-density data, the various sources of which will be addressed 
in the following subsection. 
Continuous transitions between these types are probably the reason for the
so-called {\em ``pinch effect''} 
\cite{Ker98a}, see Fig.~\ref{Fig3}(a):\par
{\noindent Upstream} of a section with
homogeneous congested traffic close to a bottleneck, there is a
so-called {\em ``pinch region''} characterized by the spontaneous 
birth of small and narrow 
density clusters, which are growing while they travel further
upstream. Wide moving jams are eventually formed by the merging or disappearance of
narrow jams, in which the cars move slower than in 
wide jams \cite{Ker98a}. Once formed, the wide jams seem to suppress the occurence of new
narrow jams in between. Similar findings were already reported by Koshi {\em et al.} \cite{KosIwOh83},
who observed that ``ripples of speed grow larger
in terms of both height and length of the waves 
as they propagate upstream''.  

\subsection{Wide scattering of congested flow-density data}\label{widescat}

The collection and evaluation of empirical freeway data is a subject with often 
underestimated problems. To make reliable conclusions, in original investigations
one should specify
(i) the measurement site and conditions (including applied control
measures),
(ii) the sampling interval,
(iii) the aggregation method,
(iv) the statistical properties
(variances, frequency distributions, correlations, survival times of traffic states, etc.),
(v) data transformations,
(vi) smoothing procedures,
and the respective dependencies on them.
\par\begin{figure}[htbp]
\begin{center}
\unitlength1cm
\begin{picture}(12,8)
\put(-0.25,4){\includegraphics[width=8.3cm]{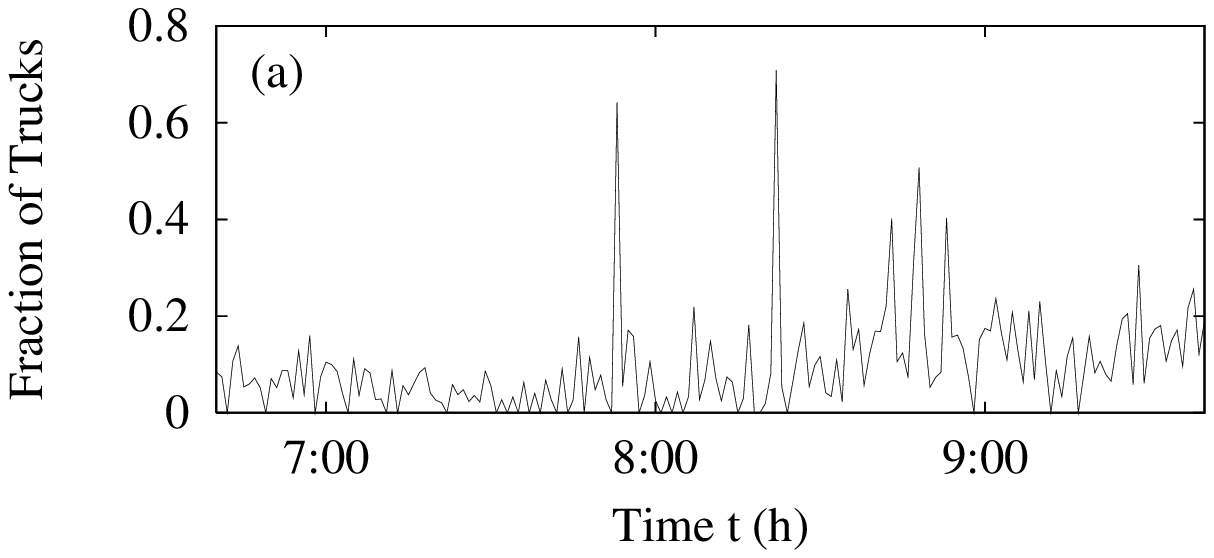}}
\put(7.9,4.3){\includegraphics[width=4.2cm]{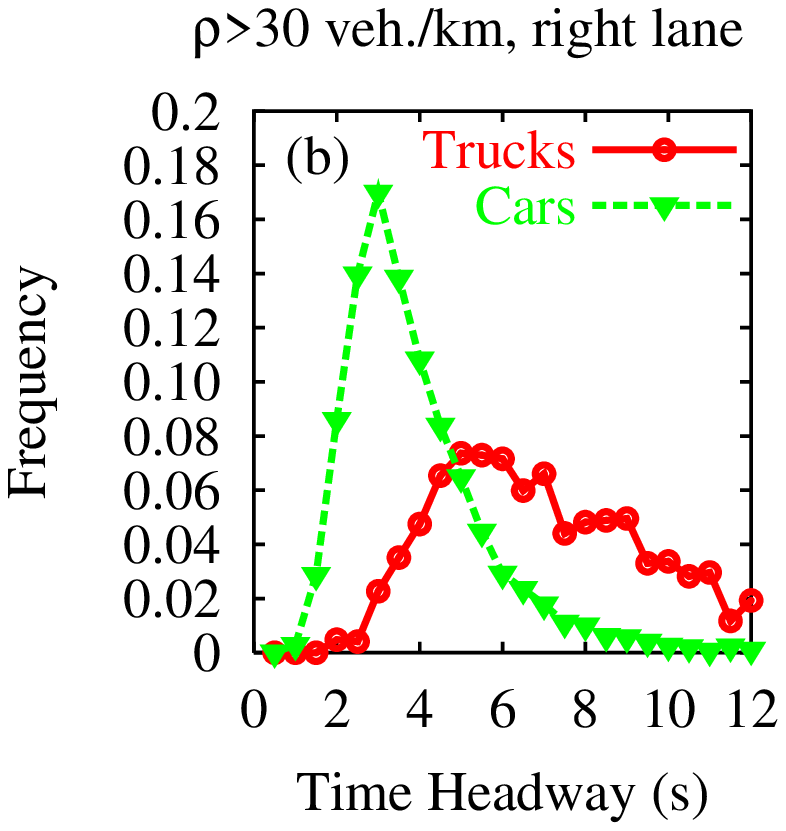}}
\put(-0.5,0.1){\includegraphics[width=4.6\unitlength]{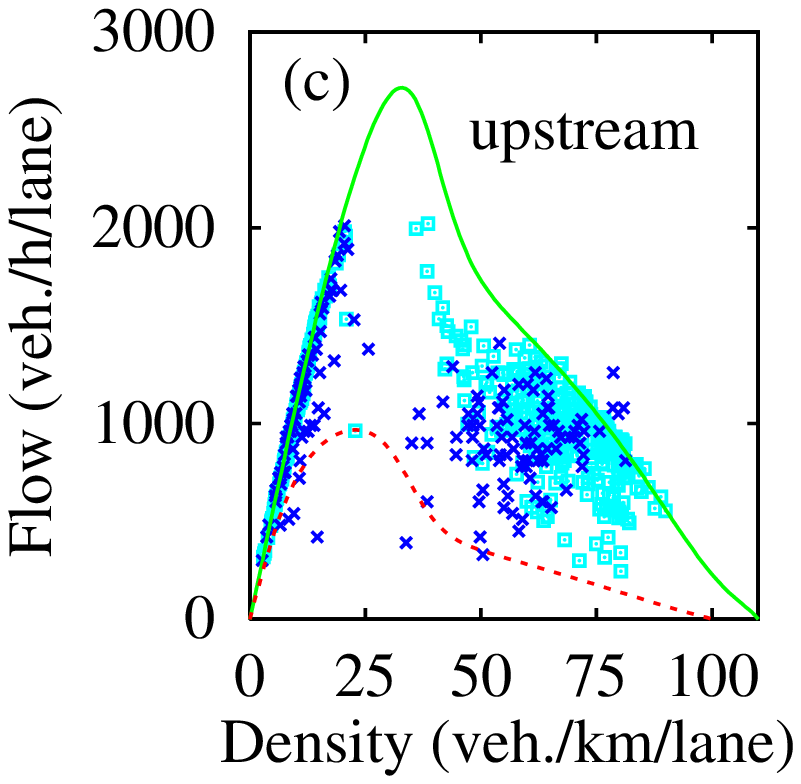}} 
\put(3.5,-0.3){\includegraphics[width=4.52\unitlength]{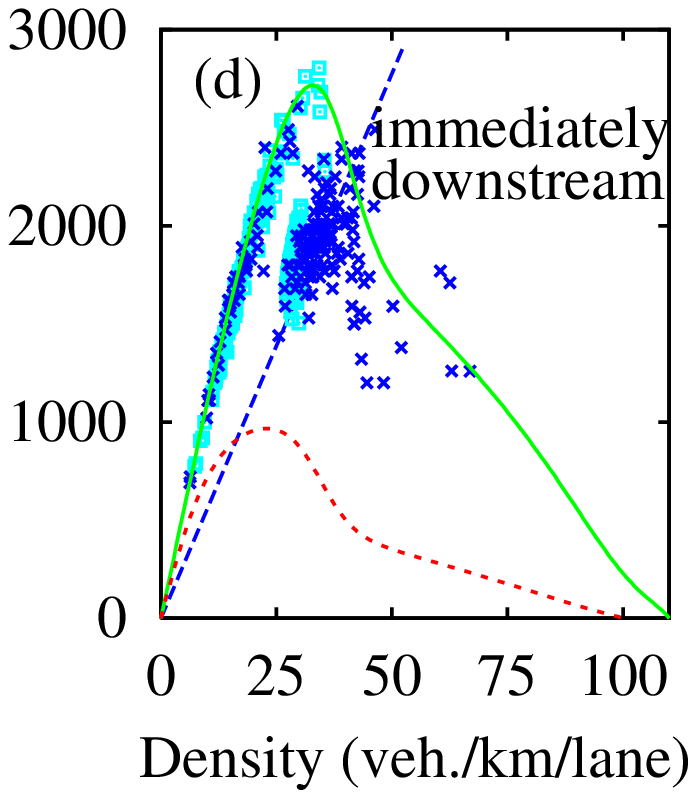}} 
\put(7.7,-0.25){\includegraphics[width=4.12\unitlength]{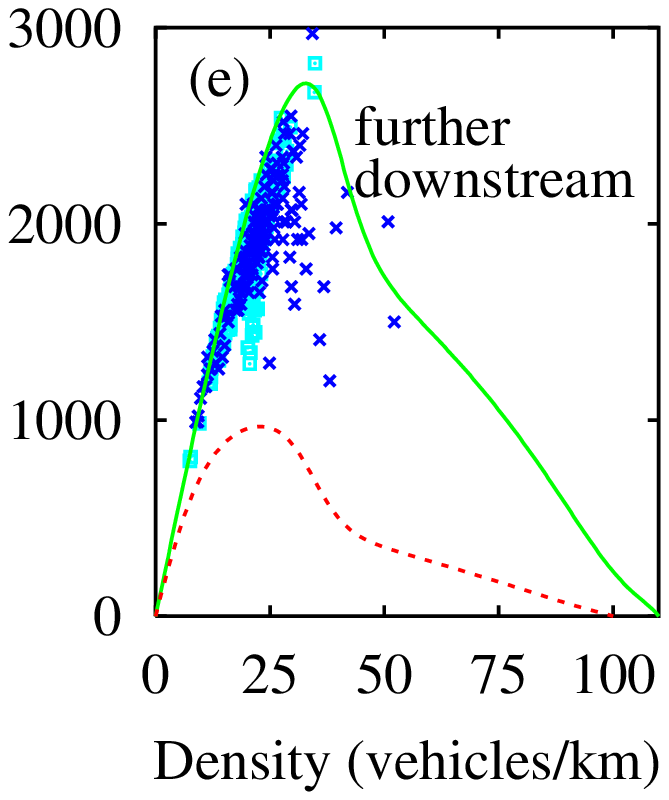}} 
\end{picture}
\end{center}
\caption[]{(a) The empirical truck fraction varies considerably in the course of time.
(b) The time headways of long vehicles (``trucks'') are on average much higher than
those of short vehicles (``cars''). (c)-(e) Assuming a fundamental diagram for cars (solid line), a separate
one for trucks (dashed line), weighting them according to the measured truck fraction, and 
using empirical boundary conditions allows to reproduce the observations in a 
(semi-)quantitatively way \cite{TreHe99a}: Free traffic (at low densities) is characterized by a
(quasi-)one-dimensional curve. (c) Data of congested traffic {\em upstream} of a bottleneck 
are widely scattered in a two-dimensional area. (d) {\em Immediately downstream} of the bottleneck,
one observes homogeneous-in-speed states reflecting recovering traffic. (e) Further
downstream the data points approach the curve describing free traffic. Dark symbols correspond
to empirical one-minute data, light ones to corresponding simulation results.\label{Fig4}}
\end{figure}
\begin{figure}[htbp]
\begin{center}
\begin{picture}(12,4.5)
\put(0,0.2){\includegraphics[width=6cm]{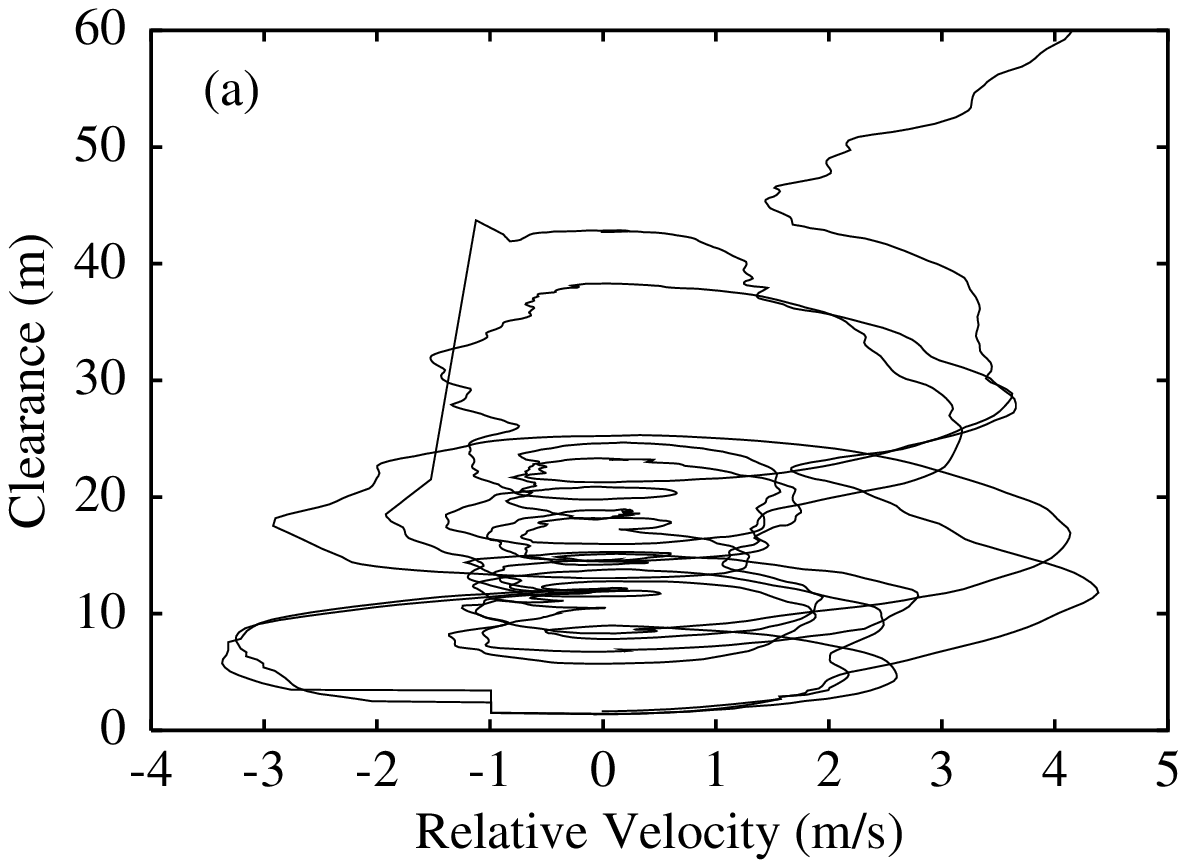}}
\put(5.9,0){\includegraphics[width=6.4cm]{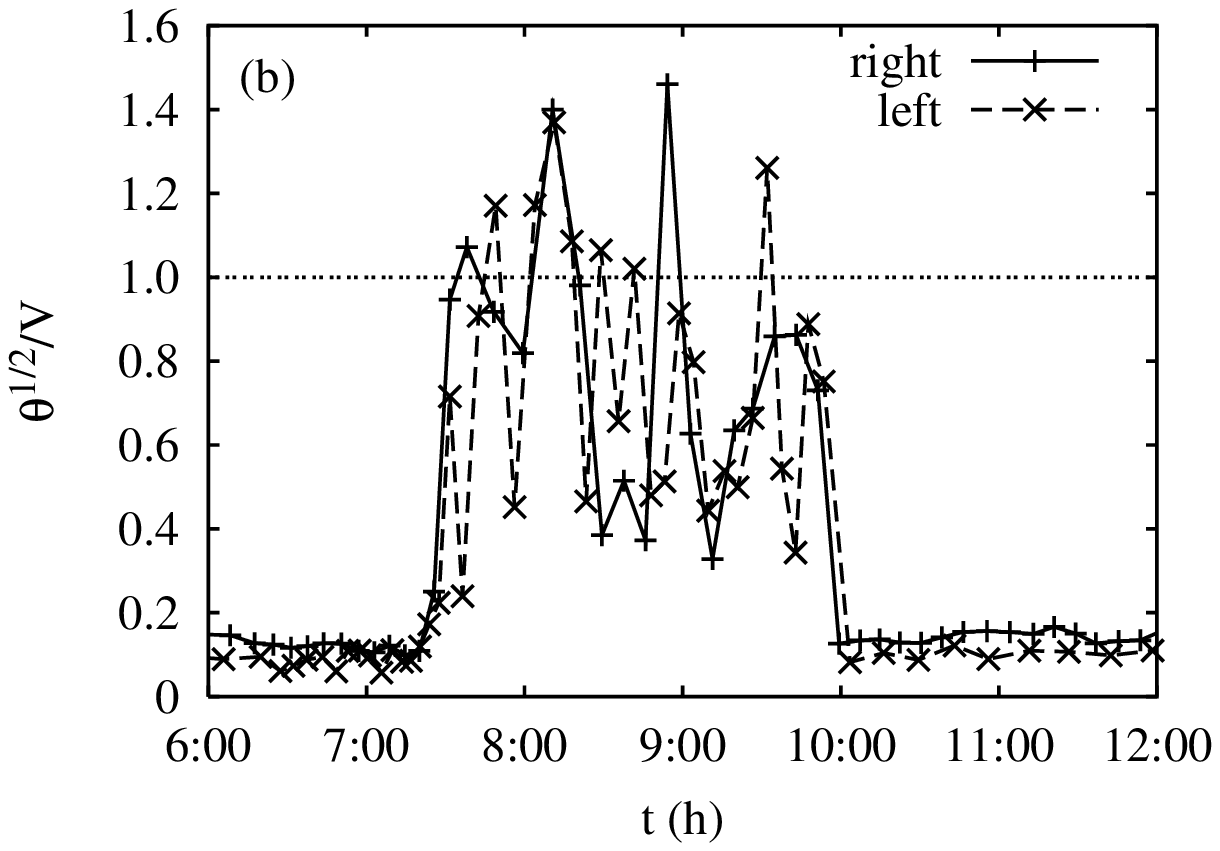}}
\end{picture}
\end{center}
\caption[]{(a) The measured oscillations of the clearance 
and the relative velocity \cite{Hoe72} 
indicate an instability in the car-following behavior \cite{HelTi98}.
(b) The empirical standard deviation $\sqrt{\theta(t)}$ of vehicle velocities
divided by the average velocity $V(t)$ is particularly large during the rush hour, where
traffic flow is congested and unstable \cite{Review,Tilch01}.\label{Fig5}}
\end{figure}

The measurement conditions include ramps and road sections with their respective
in- and outflows, 
speed limits, gradients, and curves with the respectively related capacities, 
furthermore weather conditions (like rain, ice, blinding sun, etc.), 
presence of incidents (including the opposite driving direction), 
and other irregularities such as road works, which may
trigger a breakdown of traffic flow. Moreover, one
should evaluate the long vehicles (``trucks'') separately, as their proportion varies significantly,
see Fig.~\ref{Fig4}(a).
This can explain a considerable part of the wide scattering of  congested traffic \cite{TreHe99a}, 
see Figs.~\ref{Fig4}(c), (d). Presently, this explanation is still the only one for this observation that
has been quantitatively checked with empirical data. Note that a considerable variation
of the time headways is also observed among cars, see Fig.~\ref{Fig4}(b). 
This is partly due to different driver preferences
and partly due to the instability of traffic flow, see Fig.~\ref{Fig5}(a).
While vehicle platoons with reduced time headways imply an increase
of the flow with growing density, a reduction in vehicle speeds is usually related with a 
decrease. According to Banks \cite{Bank99}, this can account for the observed erratic
changes of the flow-density data. We support this interpretation.
\par
The strong variations of traffic flows imply that all
measurements of macroscopic quantities should be complemented by 
error bars (see, e.g., Ref.~\cite{Hal1AlGu86}).  
Due to the relatively small ``particle'' numbers 
behind the determination of macroscopic quantities, the error bars are 
actually quite large. Hence, many temporal variations are within one
error bar, when traffic flow is unstable, see Fig.~\ref{Fig5}(b). 
It is, therefore, very questionable whether it is possible
to empirically prove the existence of small perturbations triggering 
a breakdown of traffic flow \cite{KerRe97} or of the ``birth'' and merging of narrow
density clusters in the ``pinch region'' \cite{Ker98a}.  At least, this would require a 
thorough statistical support. Consequently, we deny
that such kind of data are presently suited as starting point for the development of new models 
\cite{KerMic} or traffic theories \cite{Kerner3}. 
There is a considerable risk of overinterpreting particular (possibly statistical) 
features of the data recorded at special freeway sections and to construct new models
that merely simulate what has been incorporated by means of the model assumptions. 
In fact, the only reason why we believe in the correctness of these observations is the existence of
plausible deterministic traffic models reproducing these hard-to-see effects without
any special assumptions or extensions (see Fig.~\ref{Fig3} and 
Refs.~\cite{Lee3LeKi98,TreHe99b,HelTr98a}). 
\par\begin{figure}[htbp]
\begin{center}
  \includegraphics[width=7cm]{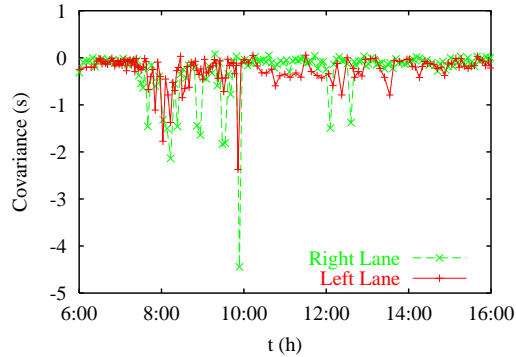}  
\end{center}
\caption[]{\label{Fig6}The covariance between headways $d_\alpha$ and inverse
  velocities $1/v_\alpha$
    shows significant deviations from zero in congested traffic, while it approximately
    vanishes in free flow, compare Fig.~\ref{Fig5}(b). 
    Even after traffic has recovered, there seem to remain 
    weak correlations between headways and vehicle speeds for a considerable time.
    These are probably a reminiscence of congestion due to platoons which have
    not fully dissolved \cite{Review,Tilch01}.}
\end{figure}
Because of the above mentioned problems, we would like to call for more refined measurement
techniques, which are required for more reliable data. 
These must take into account {\em correlations} between  different quantities, as is pointed out
by Banks \cite{Bank95}. 
\par 
For example, approximating the vehicle headways by $d_\alpha = v_\alpha \Delta t_\alpha$
(where $v_\alpha$ is the velocity and $\Delta t_\alpha$ the 
time headway of vehicle $\alpha$)
and determining arithmetic multi-vehicle
averages $\langle \dots \rangle$ at a fixed location, one obtains for the inverse vehicle flow
\begin{equation}
  \frac{1}{Q} = \langle \Delta t_\alpha \rangle 
  = \left\langle \! \frac{d_\alpha}{v_\alpha} \! \right\rangle
  = \langle  d_\alpha \rangle \left\langle \! \frac{1}{v_\alpha} 
  \! \right\rangle + \mbox{cov}\left(\!d_\alpha,\frac{1}{v_\alpha} \!\right) \, .
\end{equation}
Herein, $\mbox{cov}(d_\alpha,1/v_\alpha)$
denotes the covariance between the headways $d_\alpha$
and the inverse velocities $1/v_\alpha$, which 
is negative and particularly relevant at large vehicle densities, as expected 
(see Fig.~\ref{Fig6}). Defining the local density $\rho$ by
\begin{equation}
  \rho = 1/\left\langle d_\alpha \right\rangle
\label{proposed}
\end{equation}
and the average velocity $V$ via 
\begin{equation}
 \frac{1}{V} = \left\langle \frac{1}{v_\alpha} \right\rangle \, ,
\label{harmav}
\end{equation} 
we obtain the fluid-dynamic flow relation 
\begin{equation}
 Q = \rho V
\end{equation}
by the conventional assumption $\mbox{cov}(d_\alpha,1/v_\alpha) = 0$.  This, however,
overestimates the density systematically, since the covariance tends to be negative
due to the speed-dependent safety distance of vehicles.
In contrast, the common method of determining the density via $Q/\langle v_\alpha\rangle$
systematically underestimates the density \cite{Review,TilHe00}.
{\em Consequently, errors in the measurement of the flow and the density due to
a neglection of correlations partly account for the observed scattering
of flow-density data in the congested regime.} 
\par
A similar problem occurs when the density is determined via the time occupancy of
a certain cross section of the road. Considering that $\Delta t_\alpha 
= T_\alpha + l_\alpha/v_\alpha$,
where $T_\alpha$ is the (netto) time clearance and $l_\alpha$ 
the length of vehicle $\alpha$, we have the relation
\begin{equation}
  \rho = \rho_{\rm max} \frac{\langle l_\alpha/v_\alpha \rangle}{\langle \Delta t_\alpha \rangle}
  = \rho_{\rm max} Q \langle l_\alpha / v_\alpha \rangle = \frac{Q}{\langle l_\alpha \rangle}
 \langle l_\alpha / v_\alpha \rangle \, ,
\end{equation}
where $\rho_{\rm max} = 1/\langle l_\alpha \rangle$ is the maximum density and
$\langle l_\alpha \rangle$ the average vehicle length. For a finite detector length
$L_D$, we have to replace $l_\alpha$ by $l_\alpha+L_D$ \cite{Review,May90}. The formula $1/V =
\langle  l_\alpha / v_\alpha \rangle/\langle l_\alpha \rangle$ for the average velocity results
in the correct expression $1/V = \langle 1/v_\alpha \rangle$ only, 
if the individual vehicle lengths and velocities are not correlated, which is usually not the case.

\subsection{A quantitative theory of congested traffic states}

When Kerner started to question all traffic models with a fundamental diagram, 
physicists were used to simulate traffic in closed systems with periodic boundary conditions.
With the Kerner-Konh\"auser model, it was possible to produce
free traffic, emergent stop-and-go waves, and triggered wide moving jams \cite{Kuh84a,Kuh91a,KerKo94}. However, 
attempts to simulate ``synchronized flow'' failed even when small ramp flows were 
added to the system. They resulted in what we call a {\em pinned localized cluster}
(PLC) located at the on-ramp \cite{KerPLC} (see Figs.~\ref{Fig9} and \ref{Fig10}). 
Because of the sensitivity of the model
and problems with the treatment of open systems, it was not possible to simulate open
systems with considerable ramp flows. Other independent studies for periodic systems
with localized bottlenecks produced either homogeneous vehicle queues (HCT) 
or {\em oscillating congested traffic} (OCT) 
\cite{Lee3LeKi98,MunHsLa71,Phi77,Cre79,MakNaToMi81,ChuHu94,EmmRa95,CsaVi94,Hil95,KlaKuWe96,Naga97c}, 
but at that time nobody could make sense of these apparently inconsistent findings.
This situation changed, when Helbing {\em et al.} derived a phase diagram of
congested traffic states. They managed to simulate a macroscopic traffic model with open
boundary conditions even in extreme situations
and investigated a freeway stretch with a single ramp \cite{HelHeTr99}. 
Instead of the densities, they identified the main flow on
the freeway and the on-ramp flow as the suitable control parameters for on open system
and varied them systematically. In this way, they found that a perturbation could trigger
different kinds of congested traffic states. Moreover, the boundaries separating different states
could be related to the {\em instability diagram} for homogeneous
freeways and other characteristic quantities \cite{HelHeTr99}. For this reason, they concluded that the
phase diagram should be qualitatively the same, i.e. {\em universal}, for all microscopic,
mesoscopic, or macroscopic traffic models having the same instability diagram. This 
has been supported in the meantime \cite{HelHeTr99,TreHeHe00,Lee3LeKi99}. 
Apart from this, the phase diagram of traffic models with
different instability diagrams can be directly derived \cite{Review}. Generalizations to other kinds
of bottlenecks (e.g. gradients) have been developed as well \cite{TreHeHe00}.
\par\begin{figure}[htbp]
\begin{center}
\includegraphics[height=.9\textwidth,angle=-90]{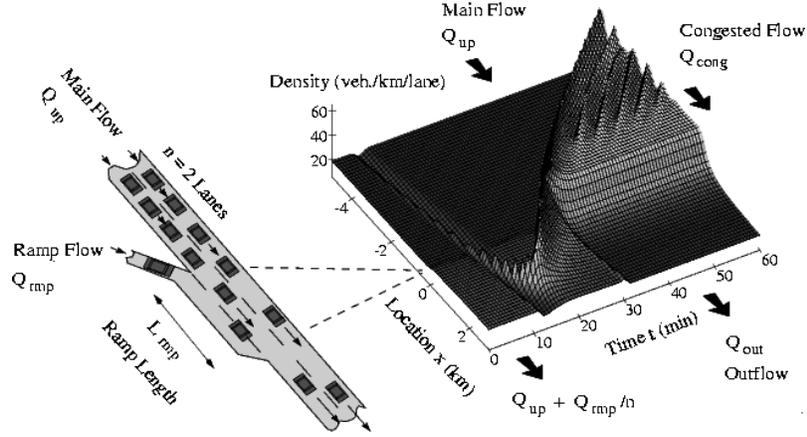}\\[3mm]
\end{center}
\caption[]{Negative perturbation triggering oscillating congested traffic.
When the traffic density has sufficiently 
increased to reach the metastable regime, the ``negative'' perturbation will 
be amplified if it only exceeds the critical amplitude. While it is small, it will move
downstream with the vehicles, so one could hope it would pass the bottleneck and leave the system.
However, when the density wave grows larger, it will reduce its speed and even change its
propagation direction.
Once it is fully developed, it moves upstream with constant velocity, since vehicles
leave the jam at the downstream front, while new ones join it at the upstream front. 
Hence, the perturbation returns to the bottleneck (see Fig.~\ref{Fig8} for this
{\em ``boomerang effect''}), and it triggers 
a breakdown of traffic, when it passes the bottleneck in upstream direction, as
it thereby reduces the effective capacity to $Q_{\rm out}$ 
\cite{Review,HelTr98a,PhysikBl}.\label{Fig7}}
\end{figure}
In the following, we will sketch the basic ideas behind the phase diagram of congested
traffic states (for a more detailed discussion see Ref.~\cite{Review}). Let us assume our
traffic model has a fundamental diagram
\begin{equation}
 Q_{\rm f}(\varrho) = \varrho V_{\rm f}(\varrho) 
\end{equation}
describing the relation between the vehicle flow $Q$, 
the {\em spatially averaged} density $\varrho$, and the average velocity $V$
in homogeneous and stationary traffic. (The flow-density relation of emergent
stop-and-go waves is characterized by a linear relation, i.e. it looks differently \cite{KerKo94}.)
Moreover, let us assume the model has ranges of stable traffic flow at small and high densities,
a range of linearly unstable traffic flow at medium densities, and ranges of metastable traffic flow
in between. This kind of instability diagram is, for example, found for the macroscopic model
used by  K\"uhne, Kerner and Konh\"auser, or Lee {\em et al.} \cite{Lee3LeKi98,Kuh84a,KerKo94}, for
the microscopic optimal velocity model \cite{BandHaNaNaShSu95}, for the non-local gas-kinetic-based
traffic model \cite{TreHeHe99}, or the microscopic intelligent driver model \cite{TreHeHe00} (among which
the first two models are rather sensitive to parameter variations, but the latter two are 
quite robust). 
\par
\begin{figure}[htbp]
\begin{center}
\includegraphics[width=.8\textwidth]{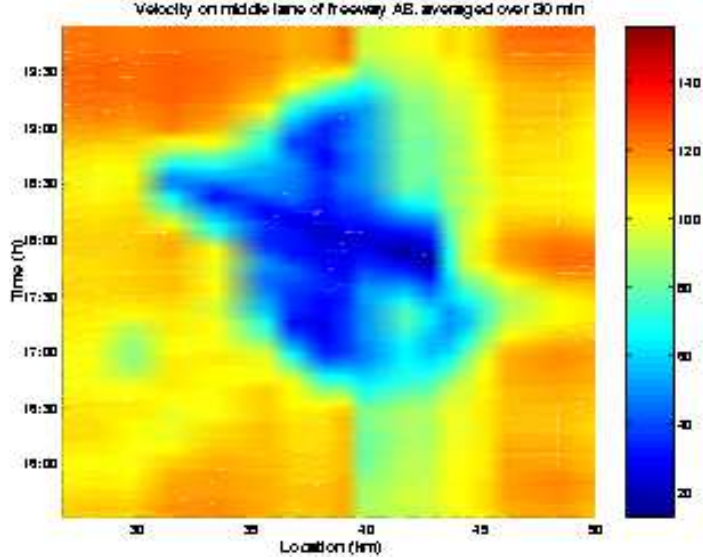}\\[3mm]
\end{center}
\caption[]{The wide moving jam left of kilometer 43 starts with a 
``boomerang effect''  and travels through the ``synchronized'' congested traffic flow 
left of kilometer 41 (dark area).
(Reproduction with kind permission of Rudolf Sollacher, Siemens AG, Munich.)
\label{Fig8}}
\end{figure}
\begin{figure}[htbp]
\begin{center}
\includegraphics[width=12cm]{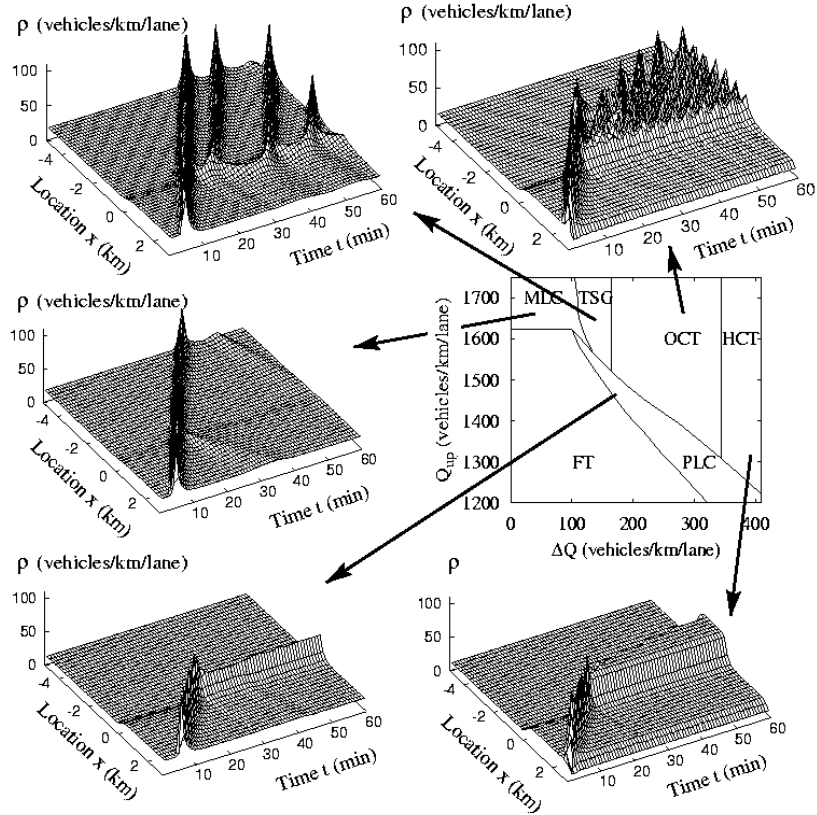}
\end{center}
\caption[]{Simulated representatives (density-over-space-and-time plots)
of the different congested traffic states, which were triggered by a big
perturbation travelling upstream. Center right: Phase diagram of the traffic states as a function
of the (upstream) traffic volume $Q_{\rm up}$ on the freeway and
the strength $\Delta Q$ of a bottleneck at location $x=0$~km, e.g. an on-ramp
with inflow $\Delta Q$ per freeway lane. 
\label{Fig9}}
\end{figure}
\begin{figure}[htbp]
\begin{picture}(16,16)
\put(0,9.5){\includegraphics[width=6cm]{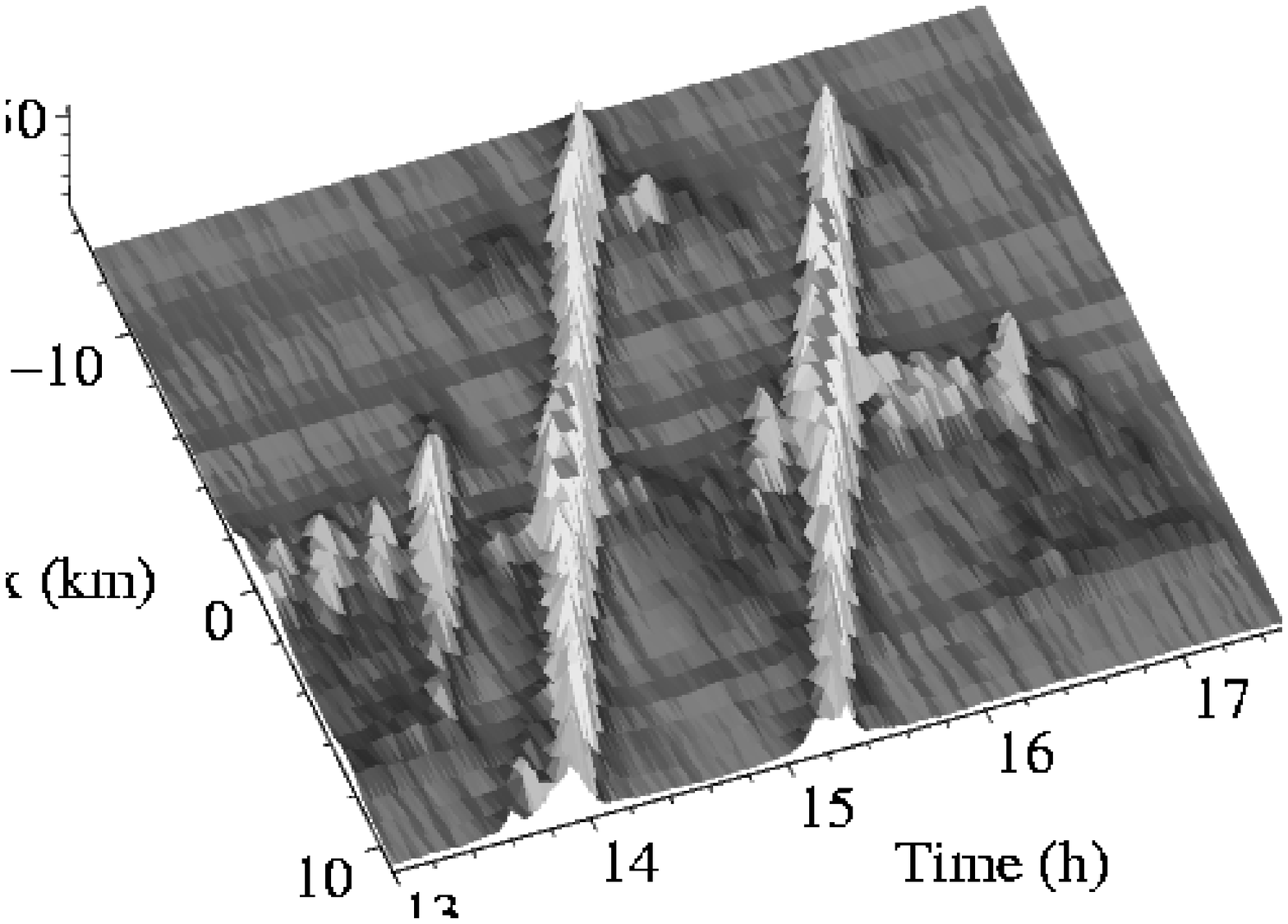}}
\put(9,8){\includegraphics[width=6cm]{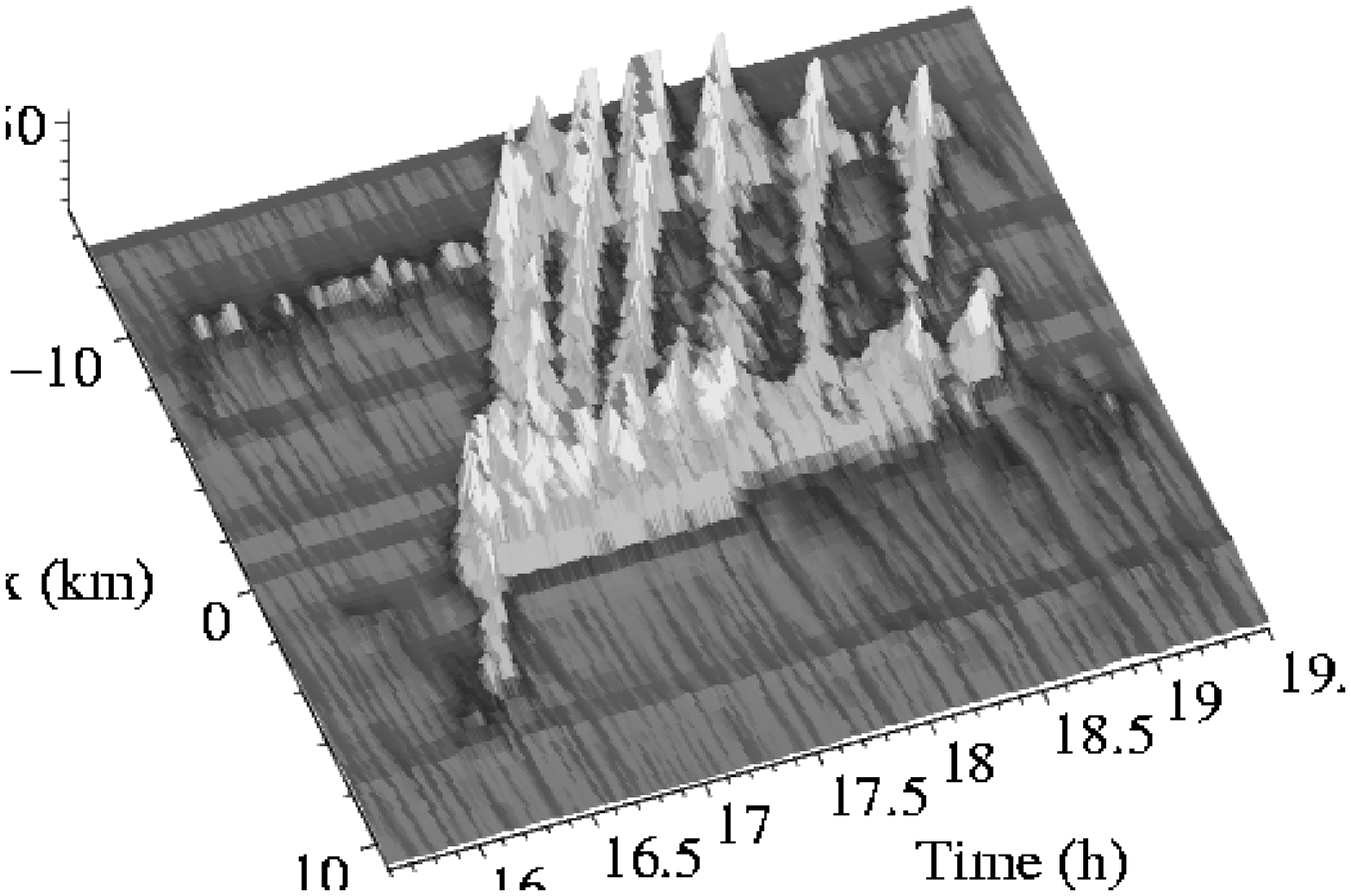}}
\put(-0.3,0.5){\includegraphics[width=6cm]{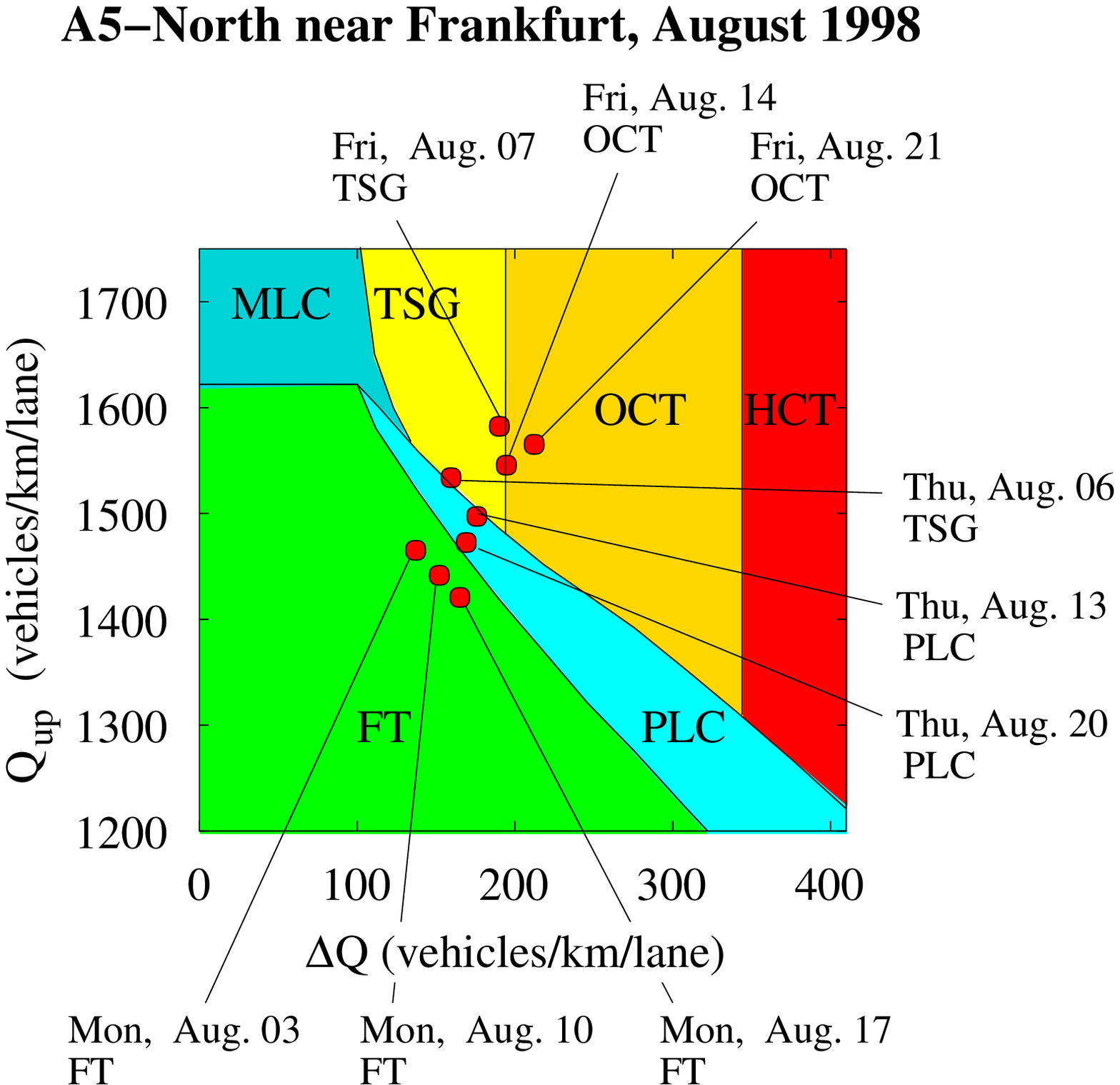}}
\put(9,0.5){\includegraphics[width=6cm]{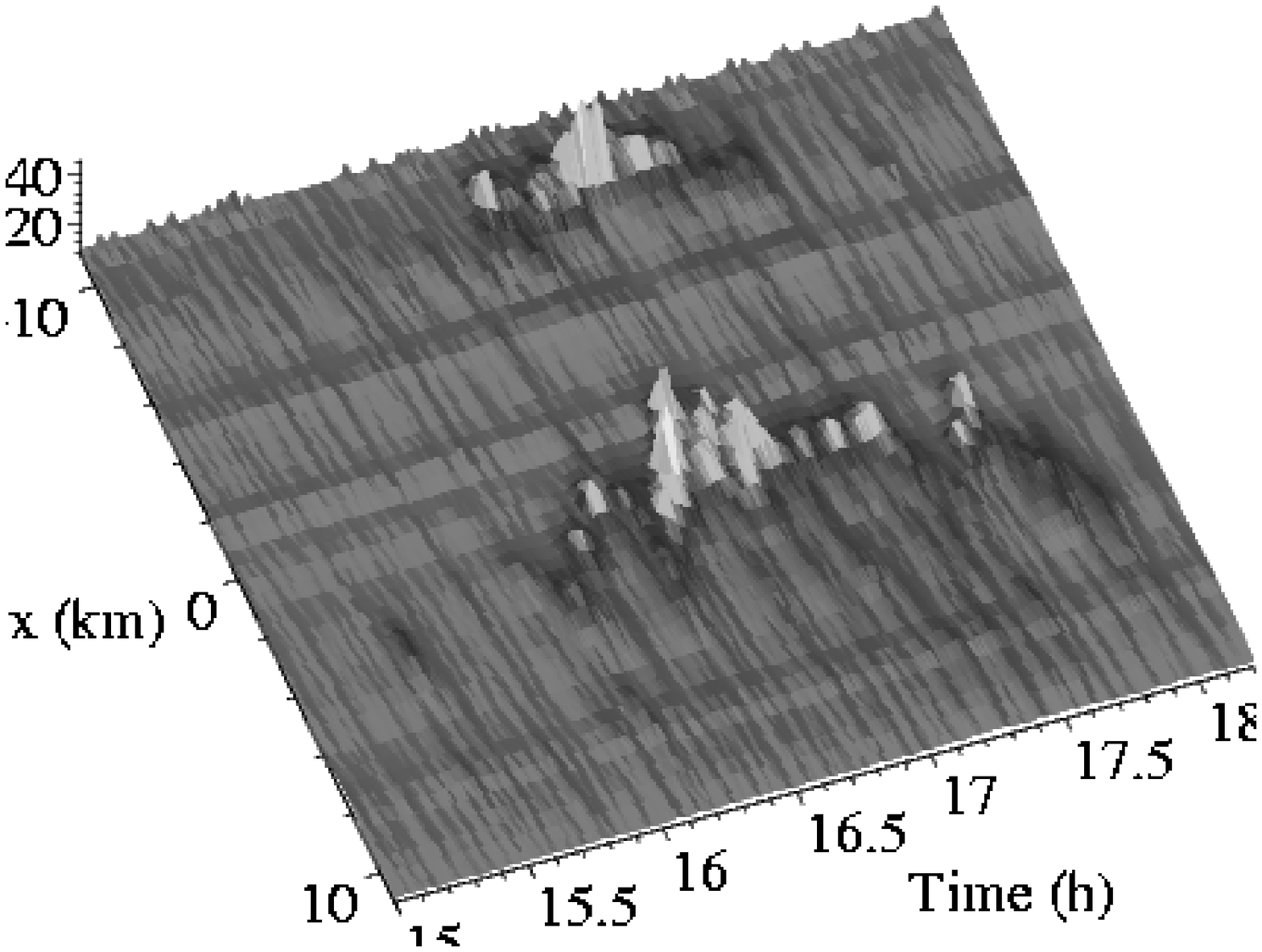}}
\put(4,15.5){\footnotesize TSG}
\put(12.5,14){\footnotesize OCT}
\put(16.3,6.5){\footnotesize PLC}
\put(17.1,4.5){\footnotesize PLC}
\end{picture}
\caption[]{Empirical examples of triggered stop-and-go traffic (TSG), oscillating
congested traffic (OCT), and pinned localized clusters (PLC), and the location of
the empirical data points in the phase diagram for the German freeway
A5 near Frankfurt.\label{Fig10}}
\end{figure}

In contrast to circular freeways, emergent ``phantom traffic jams'' are not typical for open
homogeneous freeway stretches, as it is normally impossible to reach the linearly unstable
density regime by feeding in vehicles at the upstream boundary. This is in agreement
with empirical observations \cite{Kerner4}. 
Most cases of traffic congestion on an $n$-lane freeway 
are observed upstream of on-ramps or other bottlenecks. They can be triggered by perturbations
significantly below the theoretical capacity, as soon as the sum of the upstream
freeway flow $Q_{\rm up}$ and the on-ramp flow $\Delta Q = Q_{\rm rmp}/n$ per lane exceeds the
outflow $Q_{\rm out}$ from congested traffic: If a disturbance leads to 
temporary congestion, the drivers must accelerate again and suffer some time delay,
which reduces the capacity to $Q_{\rm out}$. Therefore, the following vehicles will 
queue up, and the temporary perturbation grows to form a persistent kind of congestion.   
The initial perturbation can even be a temporary {\em reduction} of the traffic flow and/or vehicle
density, which can be caused by temporal variations of the traffic volume or even by
vehicles leaving the freeway at some off-ramp \cite{DagCaBe99,PhysikBl}, see Fig.~\ref{Fig8}. 
\par
If the total traffic volume $Q_{\rm tot} = (Q_{\rm up} + \Delta Q)$ is greater than the
dynamic capacity $Q_{\rm out}$, we will automatically end up with a growing vehicle queue
upstream of the on-ramp. The traffic flow $Q_{\rm cong}$ resulting in the congested
area normally gives, together with the inflow $\Delta Q$, the outflow $Q_{\rm out}$, i.e.
$Q_{\rm cong} = (Q_{\rm out} - \Delta Q)$ (but see the footnote on p. 1111 of Ref.~\cite{Review}
for exceptions). One can distinguish the following cases
\cite{HelHeTr99,TreHeHe00,Review} (see Fig.~\ref{Fig9}): 
If the density $\varrho_{\rm cong}$ associated with the
flow $Q_{\rm cong} = Q_{\rm f}(\varrho_{\rm cong})$ is stable, we find {\em homogeneous
congested traffic} (HCT) such as typical traffic jams during holiday seasons.
For a smaller on-ramp flow $\Delta Q$, the congested flow $Q_{\rm cong}$ is linearly unstable,
and we either find {\em oscillating congested traffic} (OCT) or {\em triggered stop-and-go traffic}
(TSG), which often emerges from a spatial sequence of homogeneous and oscillating
congested traffic (so-called {\em ``pinch effect''} \cite{Ker98a}). 
In contrast to OCT, stop-and-go
traffic is characterized by a sequence of moving jams, between which traffic flows freely.
Each traffic jam triggers the next one by inducing a small perturbation at the ramp,
which propagates downstream as long as it is small, but turns back when it has grown
large enough ({\em ``boomerang effect''}, cf.
Figs.~\ref{Fig7} to \ref{Fig10}). This, however, requires the downstream traffic flow
to be linearly unstable. If it is metastable instead (when the traffic volume is further 
reduced), a traffic jam will usually not trigger a growing perturbation. In that case, one
finds either a single {\em moving localized cluster} (MLC), or a {\em pinned localized cluster} (PLC)
at the location of the ramp. The latter requires the traffic flow in the upstream section to be stable,
so that no traffic jam can survive there. Finally, for sufficiently small traffic volumes,
we find {\em free traffic} (FT), as expected. 
\par
The different congested traffic states found
in the microsimulations (as displayed in Fig.~\ref{Fig9}) could all be identified in real traffic data
(see Fig.~\ref{Fig10} for some examples). 
Moreover, according to our first investigation results,
the traffic patterns observed on the German freeway A5 near Frankfurt 
have a typical dependence on the respective weekday and are even quantitatively
consistent with the phase diagram (see Fig.~\ref{Fig10}). 
Of course, the empirically measured patterns look less regular, as the simulation results 
displayed in Fig.~\ref{Fig9} are for a deterministic model with identical vehicle parameters.
\par\unitlength0.68cm
We will now reply to some criticism and misunderstandings:
(1) Although the phase diagram and the congested traffic states have
been derived for identical driver-vehicle units and one bottleneck only, many observations
already fit very well into this scheme. If several bottlenecks are present,
the situation becomes more complicated, but can be addressed by similar methods.
In such cases, we may find the spatial {\em coexistence} of states such as OCT and PLC, 
{\em transitions} between different states, extended congested traffic states (HCT, OCT, or TSG) 
with a {\em fixed upstream front,} and other phenomena. The phenomenon of multistability and
coexisting states is, by the way, already found for the case of 
one single bottleneck (see Ref. \cite{Lee3LeKi99} and Fig.~8 in Ref. \cite{TreHeHe00}). 
(2) Because of a certain {\em ``penetration depth''} \cite{TreHeHe00}, MLC states can propagate
through small areas of stable traffic. 
(3) The variation and scattering of the
flow-density data is well reproduced, if different driver-vehicle types are distinguished
(and  overtaking maneuvers are taken into account).
(4) The ``pinch effect'' does not correspond to a transition between different phases
(HCT, OCT, and TSG) in the course of time, but it corresponds to a certain area in
the phase diagram, where the congested flow is convectively stable but linearly unstable 
(not shown). The empirically observed increase of the vehicle velocity in ``synchronized
flow'' with the average oscillation wavelength \cite{Ker98a} is qualitatively 
well reproduced, see Fig.~\ref{Fig3}(b). (5) The flow downstream of congested traffic 
can exceed the dynamical capacity $Q_{\rm out}$, if
vehicles can enter the freeway via the ramp downstream of the congestion 
front \cite{Kerner3}. This
is practially relevant for particularly long on-ramps like freeway intersections
(see Footnote on p.~1111 of Ref.~\cite{Review}).
\section{Pedestrian Evacuation}
\unitlength0.75cm
The topics of pedestrian traffic and evacuation of buildings, stadia, and ships have recently 
attracted great interest among traffic scientists. Here, we will give a short overview only, as
there are several detailed reviews available (see Refs.~\cite{Hel97a,Review,HelMo2FaBo00,SchSh01,PED}). 
We will focus on
the {\em social-force model} of pedestrian dynamics which describes the different
competing motivations of pedestrians by separate force terms. It has the following advantages:\\
(1) The social-force model takes into account the flexible usage of space
(i.e. the compressibility of the crowd), but also the  
excluded volume and friction effects which play a role at extreme densities.
(2) The model assumptions are simple and plausible. Virtual fields \cite{BurKlaSchZi01}
or other questionable model ingredients are not necessary to obtain realistic results. 
(3) There are only a few model parameters to calibrate. 
(4) The model is robust and naturally reproduces many different observations without 
modifications of the model.
(5) Nevertheless, it is easy to consider individual 
differences in the behavior, and extensions for more complex problems such as
trail formation \cite{Hel97a,HelMo2FaBo00,HelKe1Mo97} are possible.
\par
\unitlength0.75cm
\begin{figure}
\begin{center}
\begin{picture}(12,6.8)
\put(-0.05,3.5){\includegraphics[width=12.1\unitlength]{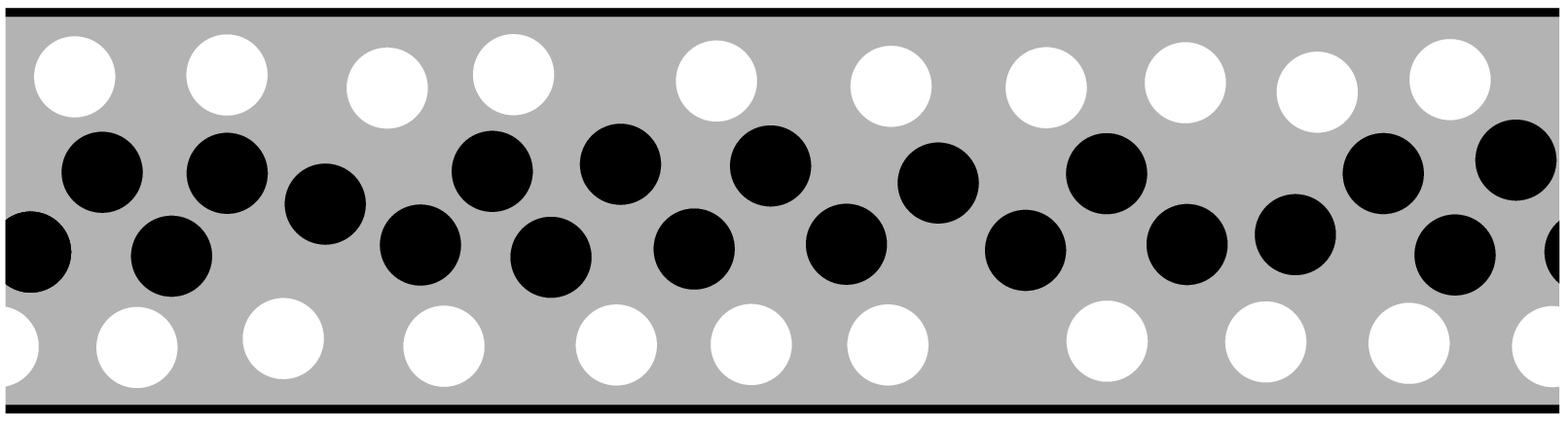}}
\put(0,0){\includegraphics[width=12\unitlength]{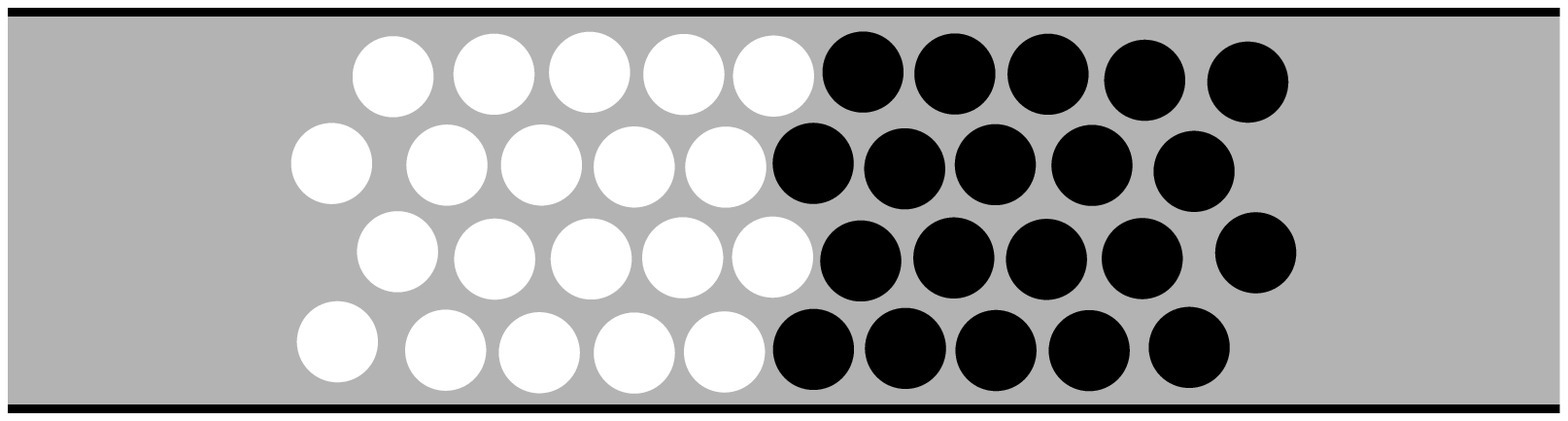}}
\put(-1,6){{\footnotesize (a)}}
\put(-1,2.5){{\footnotesize (b)}}
\end{picture}
\end{center}
\caption{(a) Formation of lanes in initially disordered pedestrian crowds 
with opposite walking directions and small noise amplitudes $\eta_i$ 
(after \cite{Review,PED,HelFaVi00a}; cf. also 
\cite{HelMo295,HelVi99}).
White disks represent pedestrians moving
from left to right, black ones move the other way round. 
(b) For sufficiently high densities and large fluctuations, we observe
the noise-induced formation of a crystallized, ``frozen'' state 
(after \cite{Review,HelFaVi00a}).\label{Fig11}}
\end{figure}
For normal, ``relaxed'' situations with small fluctuation amplitudes, 
our microsimulations of counterflows in corridors reproduce the
empirically observed {\em formation of lanes} consisting of pedestrians with the
same desired walking direction \cite{Review,HelFaVi00a,HelMo295,HelVi99}, see Fig.~\ref{Fig11}(a). 
If we do not assume periodic boundary conditions, these lanes are dynamically varying
(see the Java applet at {\tt www.helbing. org/Pedestrians/ Corridor.html}). Their number
depends on the width of the street \cite{HelMo295}, 
on pedestrian density, and on the noise level. Interestingly, one finds
a {\em noise-induced ordering} \cite{HelPl00}:
Compared to small noise amplitudes, medium ones result in a more pronounced segregation
(i.e., a smaller number of lanes). Large noise amplitudes lead to
a {\em ``freezing by heating''} effect characterized by a breakdown of ``fluid''
lanes and the emergence of ``solid'' blockages \cite{HelFaVi00a}, see Fig.~\ref{Fig11}(b).
Note that our model can explain lane formation even without assuming asymmetrical interactions
or attraction effects \cite{HelFaVi00a,HelMo295,HelVi99}. 
It is an {\em optimal self-organization} phenomenon \cite{HelVi99} 
resulting from the combination of driving and repulsive 
forces. The same model also reproduces the 
observed oscillations of the flow direction at bottlenecks \cite{Review,PhysikBl,HelMo295} 
without the need of a virtual ``floor field'' \cite{Scha01,BurKirKlaSchZi01}. {\em Cellular automaton}
Java applets from 1998 are available in the internet to visualize these phenomena 
(see {\tt www.helbing.org/Pedestrians/ Corridor.html, Door.html}). They are based on
a discretization of the social force model that can be viewed as a discrete two-dimensional
optimal velocity model \cite{Bolay98}. For other cellular automata see 
Ref.~\cite{SchSh01,BlueAd98,FukIs99a,MuraIrNa99,KluMeWaSc00}.
\par
\begin{figure}[htbp]
\begin{picture}(17,12.1)
\put(0.4,6){\includegraphics[height=6\unitlength,clip=]{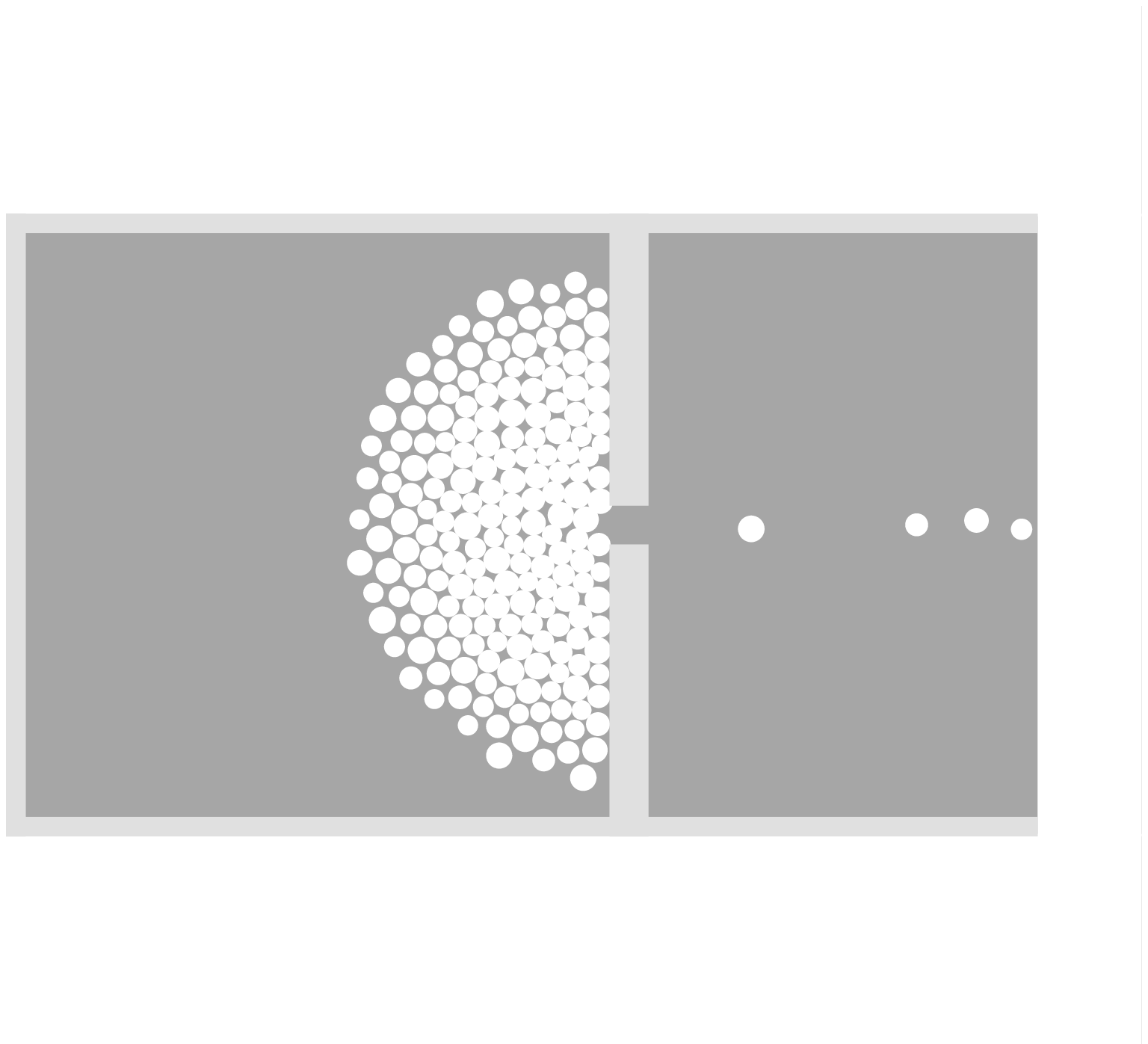}}
\put(10,11.9){\includegraphics[width=5.6\unitlength, angle=-90]{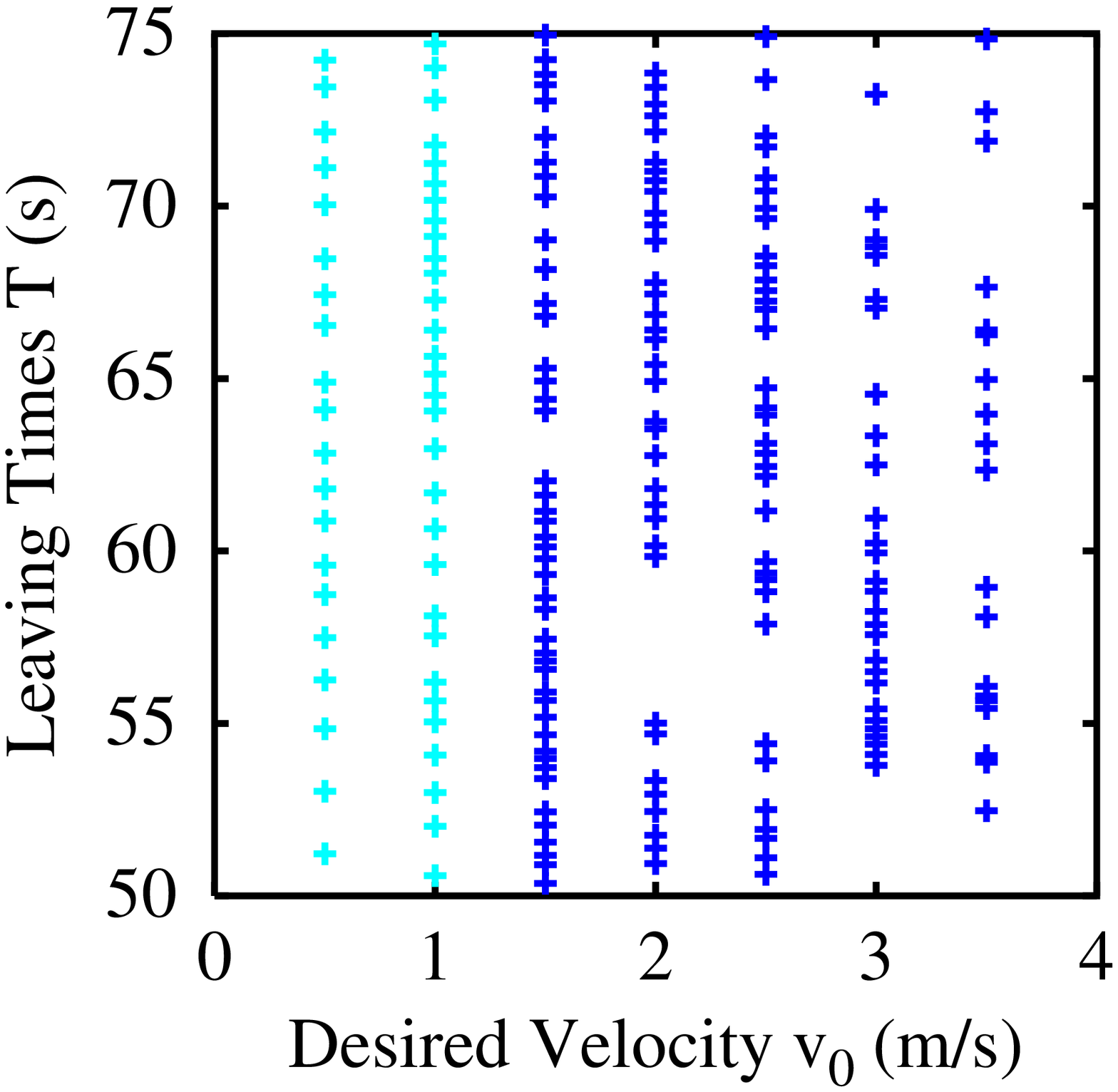}}
\put(-0.1,11.3){{\footnotesize (a)}}
\put(9.95,11.3){{\footnotesize (b)}}
\put(0.1,0.5){\includegraphics[width=7.4\unitlength,clip=]{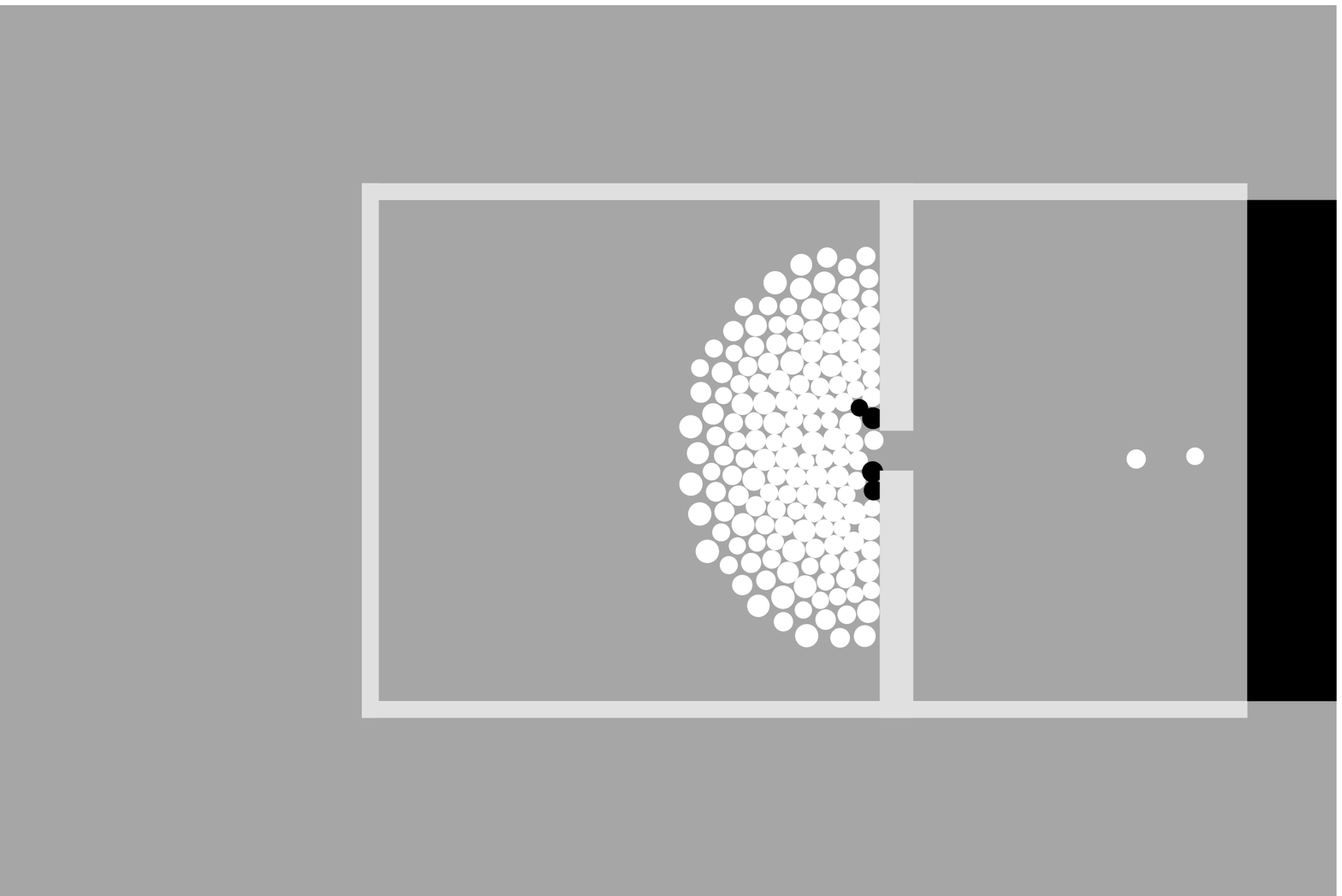}}
\put(8.1,0.5){\includegraphics[width=7.4\unitlength,clip=]{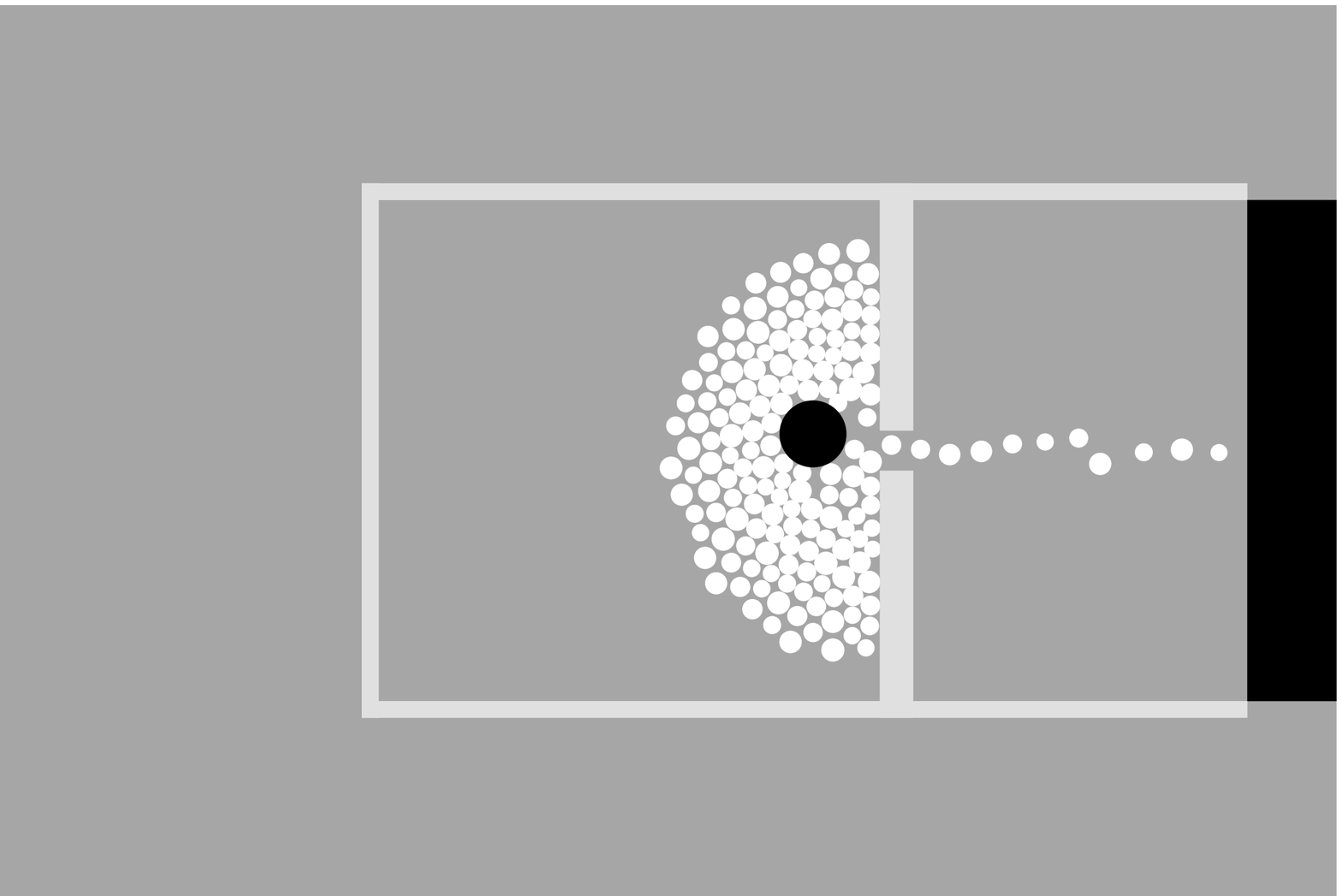}}
\put(0.1,5.5){\footnotesize (c)}
\put(8.1,5.5){\footnotesize (d)}
\end{picture}
\caption[]{(a)When the desired velocity $v_0$ is too high (e.g. in panic situations),
pedestrians come so close to each other, that their
physical contacts cause the build up of pressure and obstructing friction effects,
which results in temporary arching and clogging (for online Java simulations see
{\tt http:/$\!$/angel.elte.hu/$\tilde{\hphantom{n}}$panic/}).
(b) This is related with an
inefficient and irregular outflow, while the outflow is regular
for small enough desired velocities ($v_0 \le 1.5$~m/s) \cite{Review,PED,HelFaVi00b}.
(c) In panicking crowds, high pressures build up due to physical interactions.
This can injure people (black disks), who turn into obstacles for other pedestrians trying to leave
\cite{PED}.
(d) A column in front of the exit (large black disk) can avoid injuries by taking up pressure
from behind. It can also significantly increase the outflow \cite{PED}. In large exit areas 
used by several hundret people,
several randomly placed columns are needed to subdivide the crowd and the pressure. An asymmetric
configuration of the columns is most efficient, as it avoids equilibria of forces which may 
temporarily stop the outflow. (See {\tt http:/$\!$/angel.elte.hu/$\tilde{\hphantom{n}}$panic/}
for online Java simulations.)\label{Fig12} }
\end{figure}

``Freezing by heating'' is one of the phenomena observed in crowd stampedes. Another
one is the {\em ``faster-is-slower effect'' (or ``slower-is-faster effect'')} 
\cite{HelFaVi00b}. It can trigger a {\em ``phantom panic''} \cite{HelFaVi00b} and is caused 
by arching and clogging at bottlenecks like exits, see Fig.~\ref{Fig12}(a). 
The underlying reason is 
the friction effect occuring in dense crowds, if the desired velocity is so high that
pedestrians have physical interactions. In these situations, extreme pressure can build
up in the crowd, and people may be crushed or trampled, in this way turning into
obstacles for other fleeing people, see Fig.~\ref{Fig12}(c).
These dangerous pressures can be reduced by columns, when suitably placed
in front of the exits, see Fig.~\ref{Fig12}(d). Thereby, the number of injuries can be 
significantly reduced, and the outflows are considerably increased (see 
the Java applets at {\tt http:/$\!$/angel.elte.hu/$\tilde{\hphantom{n}}$panic/}). 
\par
\begin{figure}
\begin{center}
\begin{picture}(16,12)
\put(0,6){\includegraphics[height=5.8\unitlength,clip=]{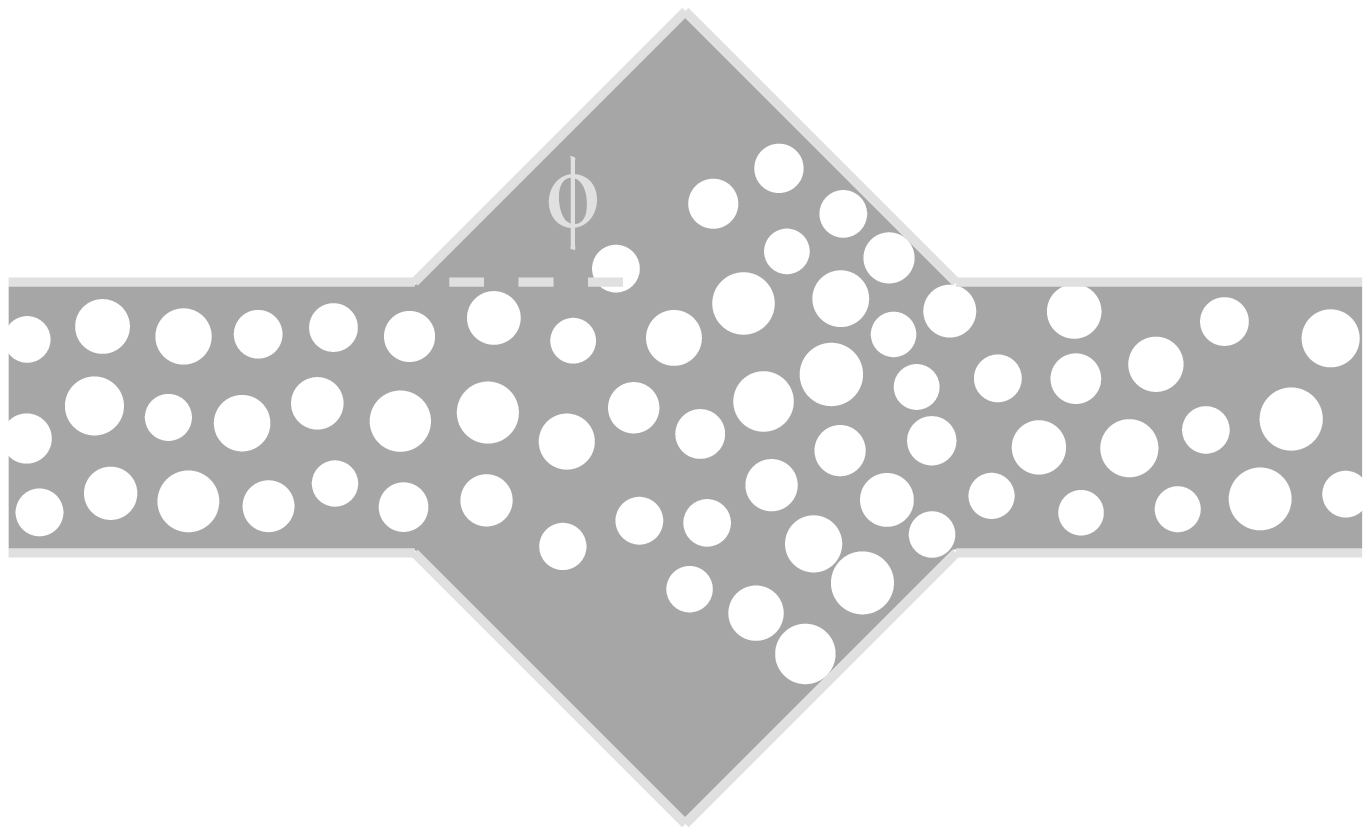}}
\put(10,11.7){\includegraphics[width=5.6\unitlength, angle=-90]{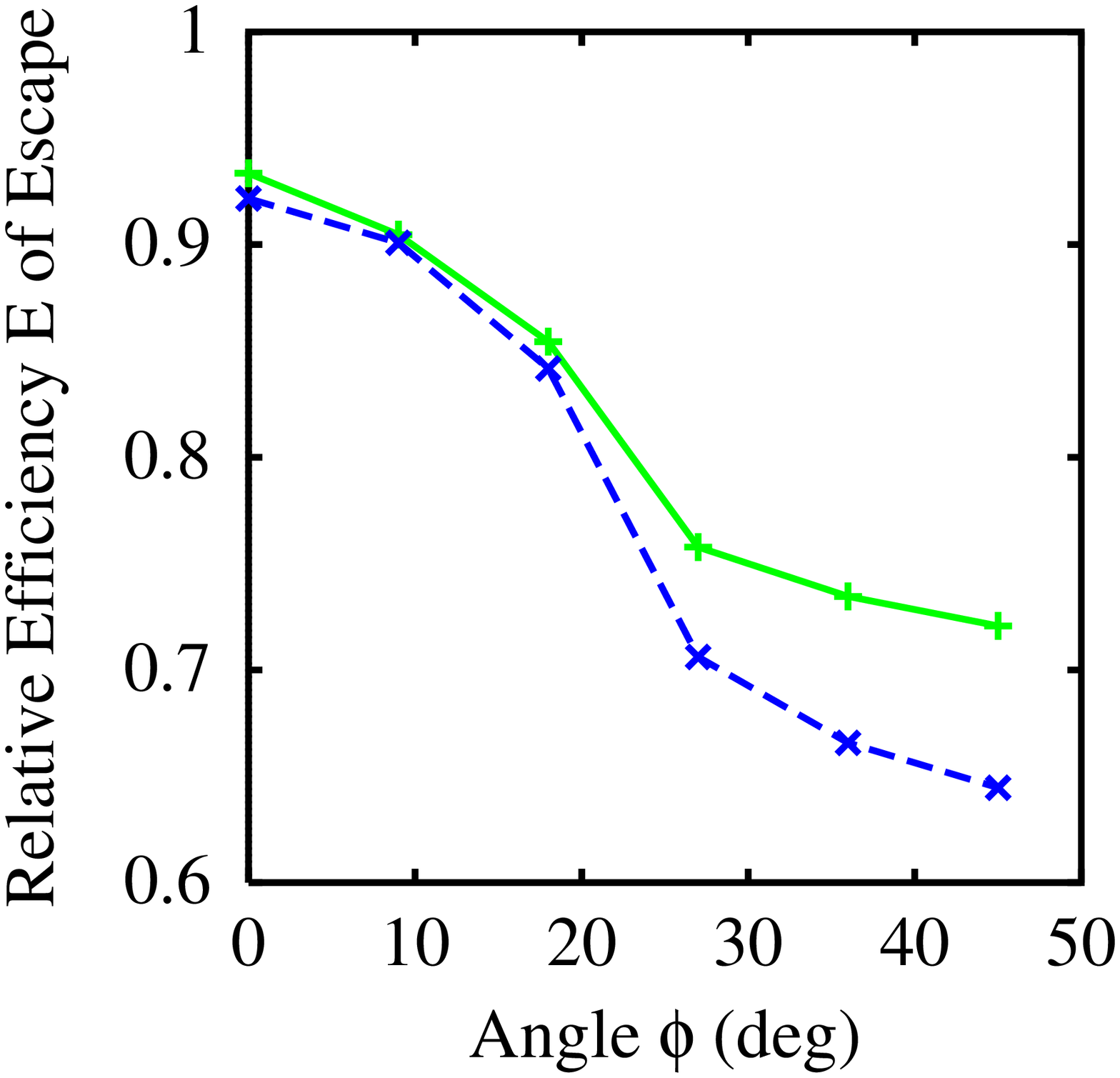}}
\put(0,10.2){\footnotesize (a)}
\put(11.6,10.7){\footnotesize (b)}
\put(-0.0,5.4){\includegraphics[height=4.35\unitlength, angle=180]{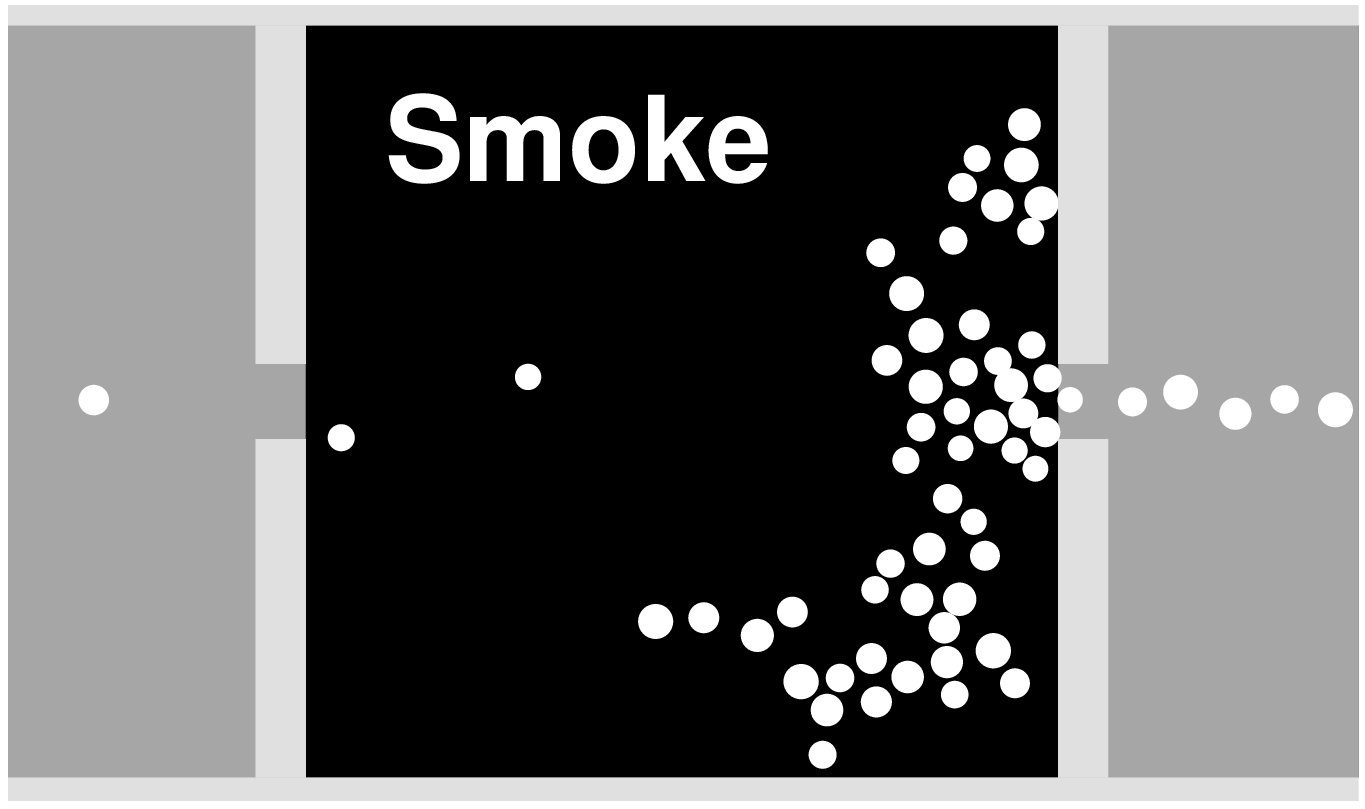}}
\put(7.9,5.7){\includegraphics[width=5.45\unitlength, angle=-90]{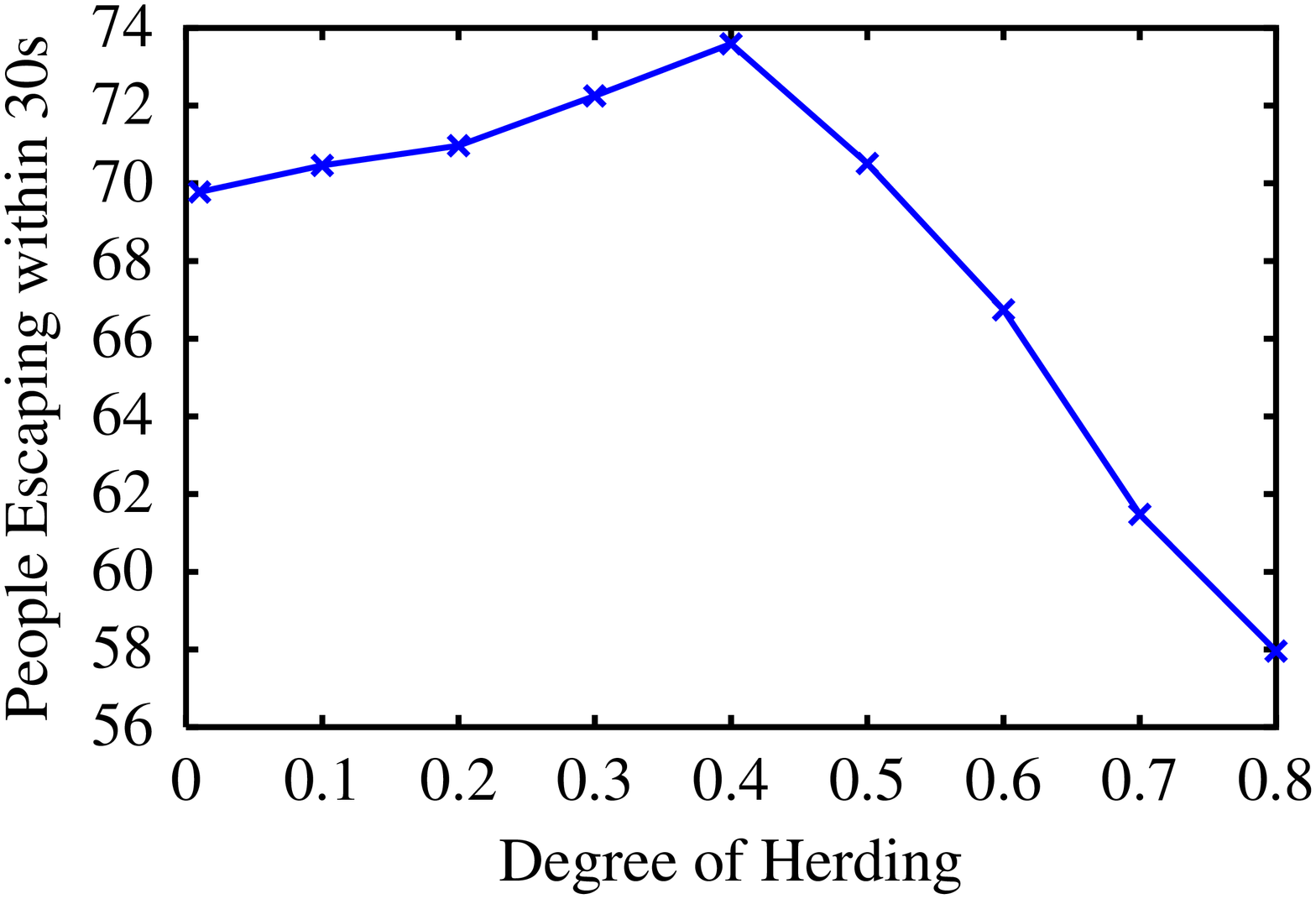}}
\put(0,5.7){\footnotesize (c)}
\put(7.85,5.7){\footnotesize (d)}
\end{picture}
\end{center}
\caption[]{(a), (b) Drop of the efficiency of leaving in corridors with widenings \cite{PED,HelFaVi00b}.
(c), (d) Herding behavior of panicking pedestrians in a smoky room (black), 
leading to an inefficient use of
available escape routes \cite{PED,HelFaVi00b}. (Online Java simulations of this effect are
provided at {\tt http:/$\!$/angel.elte.hu/$\tilde{\hphantom{n}}$panic/}.)\label{Fig13}}
\end{figure}

Not only bottlenecks are dangerous in panic situations, but also widenings \cite{HelFaVi00b}. These
can reduce the efficiency of leaving, see Figs.~\ref{Fig13}(a), (b). Another problem is
{\em herding behavior}, as it is responsible for an inefficient use of available exits \cite{Review,HelFaVi00b},
see Figs.~\ref{Fig13}(c), (d).
\par

\section{Summary and Application to Optimal Production Processes}

Nowadays, most aspects of traffic dynamics have been understood on the basis of
self-driven many-particle models. The observed phenomena can be \mbox{(semi-)}quan\-ti\-ta\-tively
reproduced by simulations using measured boundary conditions \cite{TreHeHe00}. Moreover, a
universal phase diagram of congested traffic states for freeway sections with
one bottleneck has been found, and the generalization to more complex situations
is straightforward. The reproduction of fine details, however, will require a more detailed
measurement of the interactive vehicle dynamics and the consideration of psychological
aspects. Although these may also be described in a mathematical way \cite{KerMic}, it will be hardly
possible to prove or disprove the corresponding models, i.e. the criteria demanded in
the natural sciences would have to be relaxed. A more promising research direction 
is the modelling and optimization of production processes.
For example, applying the knowledge of the ``slower-is-faster effect'' to the treatment times in
a series of production steps, we were able to increase the throughput of 
production processes in a major Chip factory by up to 39\% \cite{Fasold}.
\par
\begin{figure}[htbp]
\unitlength0.8cm
\begin{center}
\begin{picture}(14,14)
\put(0.2,14.5){\includegraphics[width=7.3\unitlength,angle=-90]{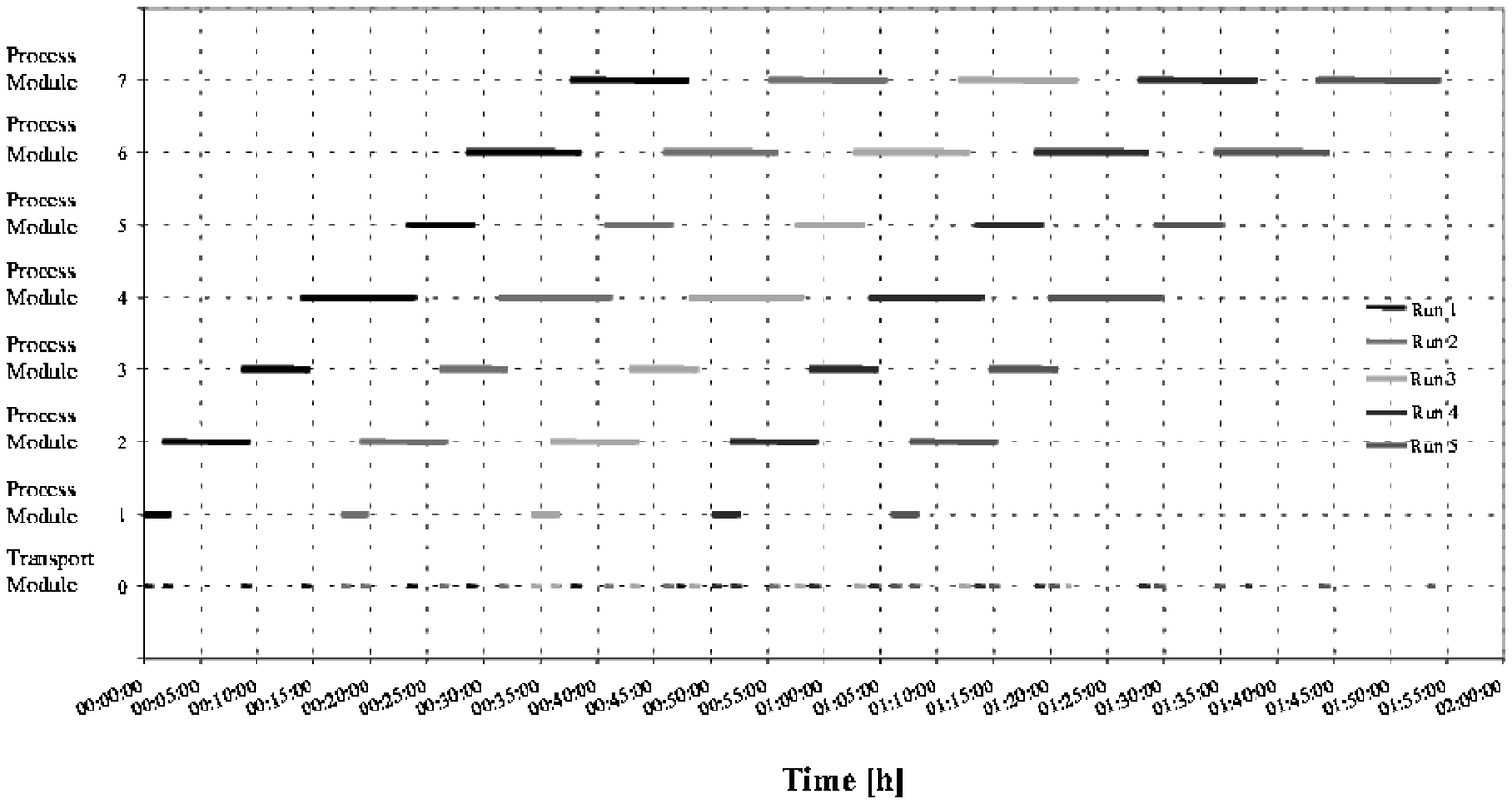}}
\put(0,7){\includegraphics[width=7\unitlength,angle=-90]{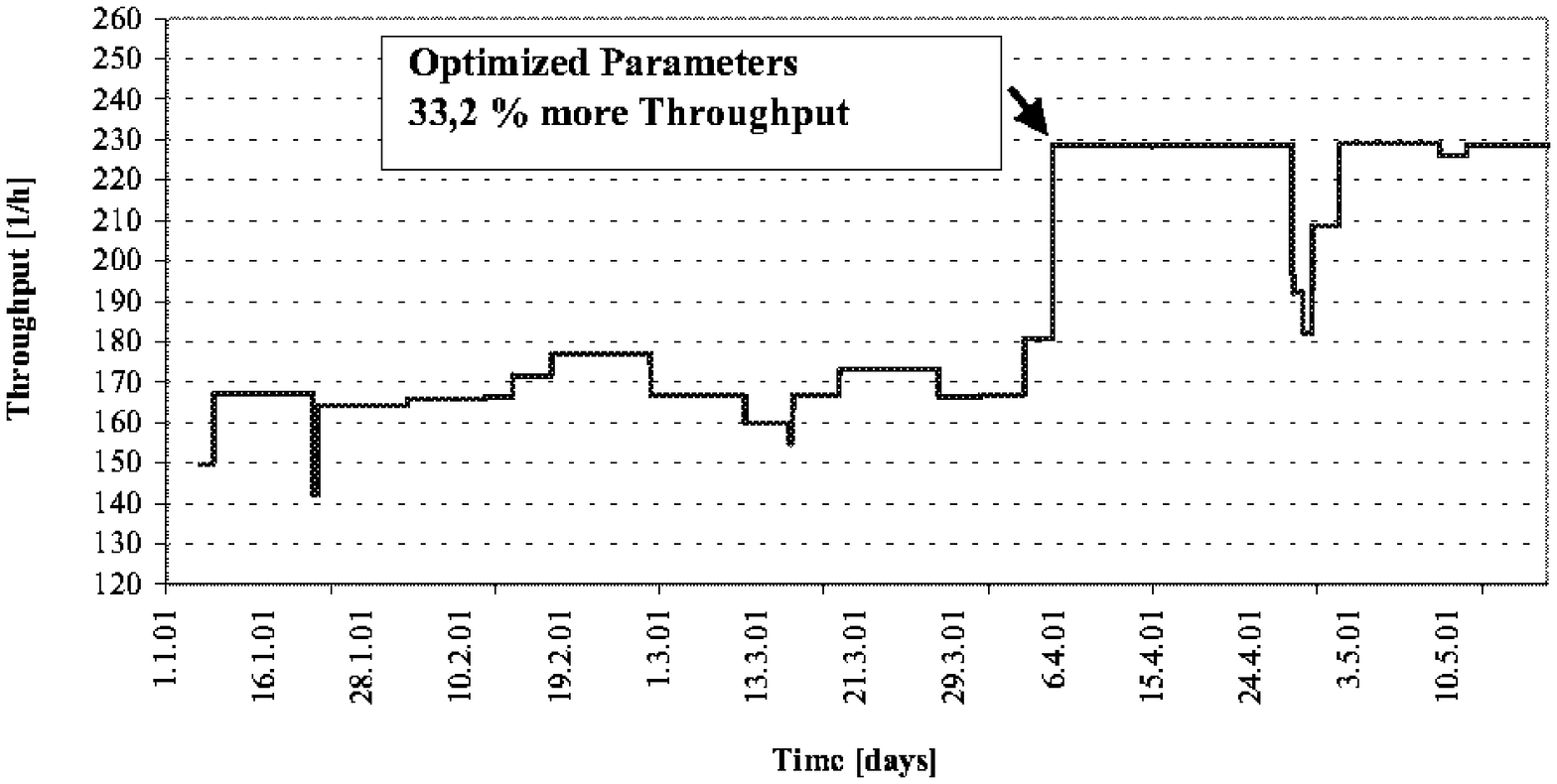}}
\put(0,14.3){\footnotesize (a)}
\put(0,6.7){\footnotesize (b)}
\end{picture}
\end{center}
\caption[]{(a) Gantt diagramm illustrating the treatment times in 
different modules, corresponding to different production steps. The limiting
factor for increasing the throughput is the transport module.
(b) Applying the knowledge of the ``slower-is-faster effect'' to the treatment times, 
we were able to increase the throughput of  
some production processes in a major Chip factory by 33\% or even 
more \cite{Fasold}.\label{Fig14}}
\end{figure}
Many conclusions from traffic models are relevant for the organization of societies,
administrations, companies, production processes,  and so on, as the basic model ingredients 
agree:
(1) The system consists of a large number of similar {\em units/entities} 
(individuals, pedestrians, cars, boxes, ...).
(2) The units are externally or internally {\em driven,} i.e. there is some energy input, e.g.,
they can move.
(3) Units are {\em non-linearly interacting,} i.e. under certain conditions,
small variations can have large effects.
The system behavior is dominated by interactions rather than by the 
boundary conditions (the external control).
(4) There is a {\em competition for resources} such as time (slots), 
space, money.
(5) Each unit has a certain {\em extension} in space or time.
(6) When units come too close, they have {\em frictional effects} and obstruct each other.

\end{document}